\documentclass[prd,aps,showpacs,floats,letterpaper,floatfix,twocolumn,
nofootinbib]{revtex4}
\usepackage{amssymb,amsmath,amsbsy}
\usepackage[dvips]{graphicx}
\usepackage{dcolumn}
\usepackage[usenames]{color}
\usepackage{euscript}
\usepackage{mathrsfs}
\usepackage{enumerate}

\begin{document}

\title{Tidal coupling of a Schwarzschild black hole and circularly
  orbiting moon}

\author{Hua Fang}
\author{Geoffrey Lovelace}
\affiliation{Theoretical Astrophysics, California
Institute of Technology, Pasadena, California 91125}

\begin{abstract}
We describe the possibility of using LISA's gravitational-wave
observations to study, with high precision, the response of a
massive central body (e.g.\ a black hole or a soliton star) to the
tidal gravitational pull of an orbiting, compact, small-mass object
(a white dwarf, neutron star, or small-mass black hole). Motivated
by this LISA application, we use first-order perturbation theory to
study tidal coupling for a special, idealized case: a Schwarzschild
black hole of mass $M$, tidally perturbed by a ``moon'' with mass
$\mu \ll M$ in a circular orbit at a radius $b\gg M$ with orbital
angular velocity \(\Omega\). We investigate the details of how the
tidal deformation of the hole gives rise to an induced quadrupole
moment ${\mathscr I}_{ij}$ in the hole's external gravitational
field at large radii, including the vicinity of the moon. In the
limit that the moon is static, we find, in Schwarzschild coordinates
and Regge-Wheeler gauge, the surprising result that {\it there is no
induced quadrupole moment}. We show that this conclusion is gauge
dependent and that \emph{the static, induced quadrupole moment for a
black hole is inherently ambiguous}, and we contrast this with an
earlier result of Suen, which gave, in a very different gauge, a
nonzero static induced quadrupole moment with a sign opposite to
what one would get for a fluid central body. For the orbiting moon
and the central Schwarzschild hole, we find (in agreement with a
recent result of Poisson) a time-varying induced quadrupole moment
that is proportional to the time derivative of the moon's tidal
field, ${\mathscr I}_{ij} = (32/45)M^6 \dot {\mathcal E}_{ij}$ and
that therefore is out of phase with the tidal field by a spatial
angle $\pi/4$ and by a temporal phase shift $\pi/2$.  This induced
quadrupole moment produces a gravitational force on the moon that
reduces its orbital energy and angular momentum at the same rate as
the moon's tidal field sends energy and angular momentum into the
hole's horizon. As a partial analog of a result derived long ago by
Hartle for a spinning hole and a static distant companion, we show
that the orbiting moon's tidal field induces a tidal bulge on the
hole's horizon, and that the rate of change of the horizon shape
(i.e. the horizon shear) {\it leads} the perturbing tidal field at
the horizon by an angle $4M\Omega$. We prefer to avoid introducing
an ingoing null geodesic, as Hartle did in his definition of the
phase shift, because the moon is in the central body's near zone
(\(b \ll 1/\Omega\)) and thus should interact with the horizon
instantaneously, not causally. We discuss the implications of these
results for LISA's future observations of tidal coupling, including
the inappropriateness of using the concepts of {\it tidal
polarizability} and {\it tidal lag or lead angle}, for the massive
central body, when discussing LISA's observations.
\end{abstract}


\date{\today}

\maketitle


\section{Introduction and Summary}
\label{sec:introduction}

\subsection{Motivations}

One of the primary scientific requirements for LISA (the Laser
Interferometer Space Antenna) is to map, in exquisite detail, the
spacetime geometries of massive black holes (and, if they exist,
other massive, compact bodies) by using the gravitational waves
emitted by inspiraling white dwarfs, neutron stars, and small-mass
black holes. This emission process has come to be called ``Extreme
Mass Ratio Inspiral'' (EMRI, pronounced emm-ree).  The possibility
of making such maps from EMRI waves was discussed by Thorne in the
early 1990s (e.g., in Refs.~\cite{Abromovici92, Thorne94}).  In 1995
Ryan~\cite{ryan1} laid the first detailed foundation for such
mapping:  he showed that, when the massive, central body is general
relativistic, axisymmetric, and reflection-symmetric, and the
orbiting object is in a near-equatorial, near-circular orbit in the
vacuum region surrounding the body, the full details of the central
body's metric are encoded in (i) the phase evolution of the waves
and also in (ii)  the evolution of the frequencies (or phases) of
wave modulation produced by orbital precession.
Phinney~\cite{Phinney00} has given the name ``bothrodesy'' to the
mapping of a black hole's metric via EMRI waves, and bothrodesy has
been identified, by the LISA International Science Team (LIST), as
one of the prime goals for LISA~\cite{Phinney02}. The initial phase
of scoping out LISA's data analysis challenges for EMRI waves is now
underway~\cite{Baracketal, Baracketal2}.

Ryan's proof that the EMRI waves carry a map of the central body's
metric ignored completely the influence of tidal coupling between
the central body and the orbiting object. Finn and
Thorne~\cite{LSFKST} have shown that, for values of the body and
object masses in the range relevant to LISA, the tidal coupling can
have an influence as large as a few percent on the evolution of the
waves' inspiral phase --- a phase that should be measurable to a
fraction of a cycle out of tens or hundreds of thousands of cycles.
Thus, the influence of the tidal coupling may be measurable with
rather high precision.  Because, in Ryan's analysis, the map is
encoded redundantly in the EMRI waves' inspiral phase and in their
modulations, it is reasonable to hope that the tidal coupling will
break that redundancy in such a manner as to permit extraction of
both the map and details of the central body's response to the tidal
gravitational pull of the orbiting object~\cite{Thorne02}.

Thorne~\cite{Thorne04} has argued that if we are to keep an open
mind about the physical nature of the central body from the outset
[e.g., if we are to allow for the possibility that it is a boson
star (e.g.~\cite{copli96,ryan96}) or a soliton star
(e.g.~\cite{lee87a}) rather than a black hole], then we must
describe the tidal coupling in a manner that can encompass all
possible types of central bodies --- a body-independent manner.

In the case of the earth and moon, the tidal coupling is normally
described in terms of the rise and fall of the earth's surface or
ocean's surface, and in terms of energy dissipation in the earth's
oceans. Noticeably different from this, the tidal coupling in the
case of a black hole has always, until now, been described in terms
of the influence of the orbiting object's gravitational field on the
hole's horizon --- the perturbation of the horizon's 2-metric
(e.g.~\cite{hartle1, hartle2}), or the conversion of the tidal field
into gravitational radiation at the horizon by gravitational
blue-shifting and the energy and angular momentum carried inward by
those waves (e.g.,~\cite{Teukolsky72}).

One tidal feature in common between a black hole, the earth, a boson
or soliton star, and all other conceivable central bodies, is the
body's tidally-induced multipole moments and multipolar
gravitational fields. It is these induced fields, acting back on the
orbiting object, that  change the object's orbital energy and
angular momentum, secularly change its orbit, and thereby alter the
emitted gravitational waves.  For this reason,
Thorne~\cite{Thorne04} has proposed that we adopt these induced
multipole fields or moments as our body-independent description of
tidal coupling when analyzing LISA data.

As a first step in exploring Thorne's proposal, we compute, in this
paper, the tidally induced quadrupole moment and its back reaction
on the orbiting object, in the special case where the central body is
a Schwarzschild black hole, and the object is a distant, circularly
orbiting moon.

\subsection{Framework and results}

Consider a moon of mass $\mu$ orbiting around a massive central body
at a large distance. When the central body is a
planet~\cite{darwin1} (see Sec.~III of Ref.~\cite{hartle2} for a
review), the external tidal field produced by the moon, denoted by
${\cal E}^{\text{ext}}_{ij}$, raises a tide on the central body and
induces a quadrupole moment ${\mathscr I}^{\text{ind}}_{ij}$ that is
proportional to ${\cal E}^{\text{ext}}_{ij}$. The proportionality
constant is the body's polarizability. Because of viscous
dissipation, the induced quadrupole moment ${\mathscr
I}^{\text{ind}}_{ij}$ will be slightly out of phase with ${\cal
E}^{\text{ext}}_{ij}$; it will have a small phase lag with respect
to the applied field. This phase lag is generally referred to as the
tidal lag angle, and can be defined equivalently as the ratio of the
tangential and radial component of the tidally-induced force acting
back on the moon. One objective of this paper is to explore whether
this type of characterization via polarizability and lag angle is
also reasonable when the central body is a black hole.

To explore this, we study a model problem where the moon is orbiting
circularly around a massive Schwarzschild black hole of mass $M$
($\gg \mu$) at large distance $b$ ($\gg M$). We assume the
separation $b$ is large enough that there exists an intermediate
region between the hole and moon where (i) gravity is weak so space
is nearly flat; (ii) the moon's tidal field does not vary
appreciably. This region is referred to as the black hole's local
asymptotic rest frame (LARF)~\cite{thorne3}. Because the spacetime
is nearly flat, one can write down the full tidal field in the LARF
(in Cartesian coordinates) to linear order in each multipole moment
as~\cite{suen2}
\begin{eqnarray}
\label{eq:tidemoments}
{\cal E}_{ij}\equiv R_{0i0j}
&=&-\sum_{\ell=0}^{\infty}
   \frac{(-1)^\ell}{\ell !}
   {\mathscr I}_{A_{\ell}}\Big(\frac{1}{r}\Big)_{,ijA_\ell}
   \nonumber \\
& & -\sum_{\ell=2}^{\infty}
   \frac{(2\ell-1)!!}{(\ell-2)!}
   {\mathscr Q}_{ijA_{\ell-2}}X_{A_{\ell-2}}.
\end{eqnarray}
Here ${\mathscr I}_{A_{\ell}}$ and ${\mathscr Q}_{A_{\ell}}$ are the
$\ell$'th internal and external moments; they are symmetric and
trace free (STF) in their tensor indices $A_\ell\equiv
a_1...a_\ell$~\cite{AL}. The ``internal moments''
$\mathscr{I}_{A_\ell}$ characterize the central body, while the
``external moments'' $\mathscr{Q}_{A_\ell}$ characterize the
gravitational fields of distant sources that perturb the central
body. In our problem, the tidal field ${\cal E}^{\text{ext}}_{ij}$
is physically the same as the external quadrupole moment ${\mathscr
Q}_{ij}$; they differ only by a constant scaling factor,
\(\mathcal{E}_{ij} = -(1/3)\mathscr{Q}_{ij}\). The internal
quadrupole moment is induced by the applied tidal field and
characterizes the tidal deformation of the central body.

Equation~(\ref{eq:tidemoments}) is the gravitational analogy to the
multipole expansion of an electromagnetic field.  It will be
sufficiently accurate for our purpose, since we shall compute the
nonspherical parts of the Riemann curvature tensor by solving the
linearized Einstein field equations. It will be shown in
Sec.~\ref{sec:setup} that only multipole moments with $\ell=0,2$ are
relevant to our problem. Dropping all other terms in
Eq.~(\ref{eq:tidemoments}) and contracting with the unit spatial
vector yields
\begin{equation}
\label{eq:tideGH}
{\cal E}_{ij} n^i n^j =
    -\frac{2M}{r^3} +
    {\cal E}^\text{ext}_{ij}n^i n^j -
    \frac{18 \mathscr{I}^\text{ind}_{ij}n^i n^j}{r^5},
\end{equation}
where we have identified $\mathscr{I}$ in Eq.~(\ref{eq:tidemoments})
as the total mass of the black hole and substituted the external
tidal field ${\cal E}^\text{ext}_{ij}$ for $\mathscr{Q}_{ij}$. In
the last term, $\mathscr{I}^\text{ind}_{ij}$ represents the
quadrupole moment induced on the black hole by the external tidal
field.

In Secs.~\ref{sec:dynamic} and~\ref{sec:rwStatic} of this paper we
compute the induced quadrupole moment in Regge-Wheeler gauge,
obtaining
\begin{equation}
\label{eq:Iind}
\mathscr{I}^{\text{ind}}_{ij} =
\frac{32}{45} M^6 \dot{\cal E}^{\text{ext}}_{ij}.
\end{equation}
The same result was recently derived by Poisson from calculating the
averaged rate of change of mass and angular momentum of the
perturbed black hole~\cite{poisson4}. Note that ${\mathscr
I}^{\text{ind}}_{ij}$ is proportional to the time derivative of
${\cal E}^\text{ext}_{ij}$ (a time derivative caused by the moon's
motion) and is therefore completely out of phase with the external
tidal field (by 90 degrees in time and 45 degrees in space). As we
will show in Sec.~\ref{sec:phase}, this out-of-phase induced moment
is gauge invariant and is responsible for the torque that changes
the orbital energy and angular momentum. Thus it is also responsible
for the tidally-induced portion of the orbital evolution and the
phase evolution of the gravitational waves.

The piece of the induced quadrupole moment that is proportional to
and in-phase with the applied tidal field is ambiguous (in a sense
that we shall discuss in Sec.~\ref{sec:gauge}); in Schwarzschild
coordinates and Regge-Wheeler gauge, it {\it vanishes}. If there had
been an unambiguous piece of ${\mathscr I}^{\text{ind}}_{ij}$ in
phase with ${\cal E}^{\text{ext}}_{ij}$, then this in-phase piece
would have defined a polarizability, and the ratio of out-of-phase
piece to the in-phase piece would have been, in a certain
well-defined sense, the small tidal lag angle. Thus, our result can
be regarded as saying that both the polarizability and the lag angle
of a black hole are ambiguous (in the sense discussed in
Sec.~\ref{sec:gauge}).

Although we find that the tidal lag angle in the LARF, in the case
of a Schwarzschild black hole, is ambiguous, we can still define and
calculate an angular tidal shift on the horizon (as opposed to in
the LARF or out at the moon). We study this horizon phase shift in
Sec.~\ref{sec:phase}. Hartle~\cite{hartle2} has calculated\footnote{We review the principal results of Hartle's investigation in Sec.~\ref{sec:hartlesummary}.} the tidal
lag angle for the problem of a bulge raised on \emph{slowly
rotating} hole's horizon by a \emph{stationary moon}, and he has
shown it to be \emph{negative}: the horizon's tidal bulge
\emph{leads} the applied tidal field due to the horizon's
teleological definition (i.e., a definition in terms of the future
fate of null rays). As in Hartle's case, we can compare the phase of
the shape of our \emph{nonrotating} horizon to our \emph{moving}
moon's position by mapping the moon to the horizon with an ingoing,
zero-angular momentum, null geodesic. In Sec.~\ref{sec:phase}, we
find that this prescription leads to a \emph{lead} angle between the
moon and the horizon
\begin{equation}
\label{eq:shift}
\delta_\text{null map} = \frac{8}{3} M \Omega + \Omega b^{*},
\end{equation}
where $\Omega$ is the orbital angular frequency of the moon and
$b^{*}$ is the moon's tortoise coordinate \(b^* \equiv b + 2M
\log(b/2M-1)\).

For comparison, Hartle's result~\cite{hartle2} for the tidal lead
angle in the case of the rotating hole and distant, stationary moon in the equatorial plane, is (after correcting a sign error, as discussed in footnote~\ref{sec:signError})
\begin{equation}\label{eq:hartlesPhase}
\delta_\text{null map}^\text{H}=\frac{2a}{3M}+\frac{a}{b}=\frac{8}{3}M\Omega_H + 4 \frac{M^2 \Omega_H}{b},
\end{equation}
Here $a$ is the hole's specific angular momentum, and $\Omega_H$ is the
horizon angular velocity. The radius of the moon's position \(b\) is sufficiently large that the moon is essentially stationary. Throughout this paper, we use the superscript ``H'' to indicate results corresponding to Hartle's system, i.e., to a system with a stationary moon and rotating horizon. Other results (without the subscript ``H'') correspond to our system of a distant moon, orbiting at frequency \(\Omega\), which perturbs a Schwarzschild black hole). 

Our result~(\ref{eq:shift}) differs from
Hartle's~(\ref{eq:hartlesPhase})---even though we initially expected
that the tidal phase shift should depend only on the difference in
angular velocities of the applied tidal field and the horizon
generators, so the results would be the same. The terms that differ arise from the particular choice to map the moon to the horizon using an ingoing, zero-angular momentum null ray.

We prefer an alternative definition of the tidal lead angle, one
that is independent of $b^*$; we prefer to define the tidal phase shift as the angle
between the perturbing tidal field at the horizon and the shear
(which is the rate of change of the shape) of the
horizon~\cite{MPBook}. This definition avoids introducing null
connections between the moon (which, at radius $b\ll\Omega^{-1}$, is
in the near zone) and the horizon. Using this definition, we find that
the shear of the central hole \emph{leads} the perturbing tidal
field at the horizon by an angle
\begin{equation}
\delta_\text{Horizon} = \delta_\text{Horizon}^\text{H} = 4 M \Omega.
\end{equation} The tidal lead angle is the same whether one considers a stationary moon perturbing a rotating hole or an orbiting moon perturbing a non-rotating hole.

The rest of paper is organized as follows. In Sec.~\ref{sec:setup},
we decompose the applied tidal field in the LARF into a
time-dependent part and a static part. In Sec.~\ref{sec:dynamic} we
analyze fully the time-dependent part and deduce the dynamical part
of the induced quadrupole moment
[Eq.~(\ref{eq:resultForInducedMoment})]. In Sec.~\ref{sec:static},
we solve for the static perturbation and discuss the ambiguity
in defining the static part of the induced quadrupole moment. In
Sec.~\ref{sec:phase}, we study the phase shift between the
deformation of the horizon and the applied tidal and compare the result with the phase shift as defined by Hartle. A brief
conclusion is made in Sec.~\ref{sec:conclusion}. Throughout the
paper, we use geometrized units with $G=c=1$.

\section{Problem Setup}
\label{sec:setup}

We study small perturbations of a non-spinning black hole caused by
an orbiting moon. The unperturbed background metric is the
Schwarzschild metric:
\begin{equation}\label{eq:bg}
ds^2 =
    -\left(1-\frac{2M}{r}\right) dt^2 + \frac{dr^2}{1-2M/r}
    + r^2 (d\theta^{2} + \sin^{2}\theta d\phi^{2}),
\end{equation}
where $M$ is the mass of the central hole. At large radii (i.e., in
the LARF), we will study the perturbations in a notation that treats
the Schwarzschild coordinates $(r,\theta,\phi)$ as though they were
flat-space spherical coordinates. These coordinates are related to
the Cartesian coordinates $(x,y,z)=(x^1,x^2,x^3)$ by
\begin{equation}
\label{eq:coord1}
(x^1,x^2,x^3)=r(\sin\theta\cos\phi,\sin\theta\sin\phi,\cos\theta).
\nonumber
\end{equation}
We will denote the radial vector with length $r$ by ${\mathbf x}$,
the unit radial vector by ${\mathbf n}$, and their components by
$x^j$ and $n^j$, respectively.

Let a moon of mass $\mu$ move along a circular orbit with radius
$b$ in the equatorial plane ($b\gg M\gg \mu$). The moon's position
is specified by
\begin{equation}
\label{eq:orbit}
{\mathbf x}^\text{s} = b\ {\mathbf n}^\text{s}= b\
(\cos{\Omega t}, \sin{\Omega t}, 0),
\end{equation}
where the superscript ``s" stands for the ``source'' of the
perturbation and $\Omega=\sqrt{M/b^3}$ is the moon's orbital angular
frequency, satisfying $\Omega b\ll1$. The moon's tidal field
\(\mathcal{E}_{ij}^{\text{ext}}\) is the double gradient of the
moon's Newtonian gravitational potential. Its value in the LARF (at
\(r\ll b\) but \(r\gg M\)) is well approximated by
\begin{eqnarray}
\label{eq:tideext}
{\cal E}^{\text{ext}}_{ij}
&=&
    -\left.\left(\frac{\mu}{|{\mathbf x}-{\mathbf x}^\text{s}|}
    \right)_{,ij}\right|_{r=0} \nonumber \\
&=& \frac{\mu}{b^3}
    \left(\delta_{ij}-3 n^\text{s}_i n^\text{s}_j\right).
\end{eqnarray}
Note that although the applied tidal field is defined in the LARF,
the induced quadrupolar field \(\mathscr{I}^\text{ind}_{ij}\) of
greatest interest is not in the LARF, but further out in the
vicinity of the moon's orbit, where it interacts with the moon.

The tidal field~(\ref{eq:tideext}) can be decomposed into spherical,
harmonic modes~\cite{thorne1}. The result of the decomposition is
\begin{eqnarray}
\label{eq:tide2m}
{\cal E}^{\text{ext}}_{ij}
&=&
  \frac{\mu}{b^3}\sqrt{\frac{6\pi}{5}}
    \left(
      \sqrt{\frac{2}{3}}{\mathscr{Y}}^{20}_{ij}
      - {\mathscr{Y}}^{22}_{ij} e^{-i\omega t}
      - {\mathscr{Y}}^{\text{2--2}}_{ij}e^{ i\omega t}
    \right)\nonumber\\
&\equiv&
  {\cal E}^{\text{ext},20}_{ij}+{\cal E}^{\text{ext},22}_{ij}
  +{\cal E}^{\text{ext},\text{2--2}}_{ij}
\end{eqnarray}
with $\omega\equiv2\Omega$ and ${\cal E}^{\text{ext},2m}_{ij}$
($m=0,\pm 2$) equal to the corresponding $\mathscr{Y}^{2m}_{ij}$
term. Here the $\mathscr{Y}^{2 m}_{ij}$ are position-independent,
rank-2, symmetric trace-free (STF) tensors defined in
Eqs.~(\ref{eq:y20tensor})--(\ref{eq:y22tensor}) and are related to
the familiar $\ell=2$ spherical harmonics $Y^{2m}(\theta,\phi)$ by
Eq.~(\ref{eq:stfconvert}). (See Eqs.~(2.7)--(2.14) of
Ref.~\cite{thorne1} for the general mapping between order $\ell$
spherical harmonics and rank-$\ell$ STF tensors). The explicit
values of the tidal field components are
\begin{subequations}
  \begin{eqnarray}
{\cal E}^{\text{ext},20}_{ij}\label{eq:staticExtTide}
    &=&
       -\frac{\mu}{2b^3}\left(
       \begin{array}{*{3}{c}}
     1 & 0 & 0\\
     0 & 1 & 0\\
     0 & 0 & -2
       \end{array}\right), \label{eq:tide20}\\
{\cal E}^{\text{ext},2\pm2}_{ij}
    &=& \label{eq:tide22}
       -\frac{3\mu}{4b^3}\left(
       \begin{array}{*{3}{c}}
     1 & \pm i & 0\\
     \pm i & -1 & 0\\
     0 & 0 & 0
       \end{array}\right)
       e^{\mp i\omega t}.
  \end{eqnarray}
\end{subequations}

The tidal field ${\cal E}_{ij}^{\text{ext}}$ [Eq.~(\ref{eq:tide2m})]
is the source of perturbations of the central hole; it is an
even-parity $\ell=2$ external tidal field. We shall therefore
perform our calculation in the even-parity Regge-Wheeler gauge, mode
by mode ($\ell=2,m=0,\pm2$). The tidal field ${\cal
E}_{ij}^{\text{ext}}$ also sets the outer boundary condition for the
problem: the $O(r^0)$ terms in the perturbed tidal field ${\cal
E}_{ij}$ must go to ${\cal E}^{\text{ext}}_{ij}$ in the LARF
[Eq.~(\ref{eq:tideGH})].

The inner boundary condition is set differently, depending on
whether the perturbations are static or time-dependent. For the
static perturbations generated by ${\cal E}^{20}_{ij}$, we impose a
``regularity boundary condition": the perturbations must be
physically finite at $r=2M$. For the time-dependent perturbations
generated by \(\mathcal{E}_{ij}^{\text{ext},2\pm 2}\), we impose the
``ingoing-wave boundary condition": the perturbations have the
asymptotic behavior $\sim e^{\mp i\omega r^*}$ when approaching the
horizon. Here $r^*$ is the tortoise coordinate \(r^*\equiv r + 2M
\log(r/2M-1)\).

\section{Time-Dependent Part of the Perturbation}
\label{sec:dynamic}
\subsection{The perturbed metric}\label{sec:tdmoment}

We will specialize to $(\ell,m)=(2,2)$ in solving for the
time-dependent part of the metric perturbation. The
$(\ell,m)=(2,-2)$ results can be obtained by complex conjugating the
$(2,2)$ results. For briefness, a superscript ``$22$'' will not be
added to quantities calculated in this harmonic mode in this
section, unless a distinction is needed. Throughout this section, we
refer to Appendix~\ref{sec:pt} for details of the perturbation
calculation.

In the standard Regge-Wheeler gauge, the $(\ell,m)=(2,2)$
time-dependent perturbations take the form~\cite{regge1}
\begin{eqnarray}\label{eq:mp22}
h^{(22)}_{ab} &=&
    Y^{2 2}(\theta,\phi) e^{-i\omega t} \nonumber \\
&\times&
    \left|
    \begin{array}{*{4}{c}}
     H(1-\frac{2M}{r}) & H_1 & 0 & 0\\
     H_1 & H(1-\frac{2M}{r})^{-1} & 0 & 0\\
     0 & 0 & r^2 K & 0 \\
     0 & 0 & 0 & r^2 K\sin^2\theta
    \end{array}\right|. \nonumber\\
\end{eqnarray}
Here $H$, $H_{1}$ and $K$ are functions of $r$ alone. These radial
functions are solutions of the perturbed Einstein equations; they
can be constructed from the Zerilli function ${\mathbb Z}(r)$
\cite{zerilli1}, which satisfies a second order ordinary
differential equation [Eq.~(\ref{eq:zerilli})]. Specifically, $H_1$,
$K$ and $H$ are given in terms of ${\mathbb Z}(r)$ by
Eqs.~(\ref{eq:H1Zerilli})--(\ref{eq:HfromZ}). Instead of solving for
${\mathbb Z}$(r) directly, one may obtain the Zerilli function from
its odd-parity correspondent, the Regge-Wheeler function ${\mathbb
X}(r)$, which obeys a simpler differential
equation~\cite{regge1,edelstein1} that is easier to solve
[Eq.~(\ref{eq:rw})]:
\begin{equation}
\left[\frac{d^{2}}{dr^{*2}} + \omega^2 -
\Big(1-\frac{2M}{r}\Big)
\Big(\frac{\ell(\ell+1)}{r^2}-\frac{6M}{r^3}\Big)\right]
\mathbb{X}(r)=0,\nonumber
\end{equation}
where $d/dr^*=(1-2M/r)d/dr$. The Zerilli function ${\mathbb Z}(r)$
is expressed in terms of ${\mathbb X}(r)$ by Eq.~(\ref{eq:Chandra}).
Thus, the metric perturbation is determined by the single radial
function $\mathbb{X}(r)$, by way of Eq.~(\ref{eq:Chandra}) to get
\({\mathbb Z}(r)\) and then
Eqs.~(\ref{eq:H1Zerilli})--(\ref{eq:HfromZ}) to get $H_1$, $K$, and
$H$.

The analytic solution for ${\mathbb X}$(r) with the ingoing-wave
boundary condition at horizon was derived by Poisson and Sasaki
\cite{poisson3}. Their solution, $X^H$ in their notation and for the
limiting case $\omega r\ll 1$, is what we have used in our analysis.
With our slow motion assumption $\Omega b\ll 1$, $X^H(\omega r\ll
1)$ will be sufficient to cover the region inside the moon's
orbit---including the LARF, where we read out the induced quadrupole
moment. Following Poisson and Sasaki's notation, we define the
dimensionless quantity
\begin{equation}
\varepsilon\equiv 2M\omega.
\end{equation}
We then combine Eqs.~(3.4), (3.11), and (3.12) of
Ref.~\cite{poisson3} to obtain
\begin{eqnarray}
\label{eq:scaleFactor}
X^H(\omega r\ll 1)\label{eq:xh}
&=& {\EuScript A}\left(\frac{r}{2M}\right)^3
    e^{i\omega(r-2M)} \nonumber \\
&\times& F(c_1,c_2;c_3;1-\frac{r}{2M}) e^{-i\omega r^*},
\end{eqnarray}
where ${\EuScript A}$ is an overall scaling factor that did not
appear in Ref.~\cite{poisson3} but will be determined by the outer
boundary condition in our problem; $F$ is the hypergeometric
function with parameters [Eq.~(3.11) of Ref.~\cite{poisson3} with
$\ell=2$]
\begin{eqnarray}
c_1 &=& -i\varepsilon + O(\varepsilon^2), \nonumber \\
c_2 &=& 5-i\varepsilon + O(\varepsilon^2), \\
c_3 &=& 1-2i\varepsilon. \nonumber
\end{eqnarray}
Note that expression (\ref{eq:xh}) for $X^H$ is only accurate to
first order in $\varepsilon$. We then expand Eq.~(\ref{eq:xh}) in
large $r$ and keep terms to first order in $\varepsilon$
\begin{equation}\label{eq:xh1epsilon}
X^H =
    \EuScript{A}\left[
        \Big(1+\frac{13}{12}i\varepsilon\Big)\tilde{r}^{3}
      + \sum_{n=5}^{\infty} \frac{i\varepsilon}{n\tilde{r}^{n-3}} +
        O(\varepsilon^2)\right],
\end{equation}
where $\tilde{r}\equiv r/2M$ is the dimensionless radius. Next, we
use Eq.~(\ref{eq:Chandra}) to get ${\mathbb Z}(r)$. Then the
perturbed metric components can be constructed using
Eqs.~(\ref{eq:H1Zerilli})--(\ref{eq:HfromZ}). In the following all
quantities will be calculated up to first order in $\varepsilon$ and
we will suppress ``$O(\varepsilon^2)$" in our expressions.

\subsection{Induced quadrupole moment in the LARF}

Once the perturbed metric is known, it is straightforward to
calculate the full Riemann tensor and extract from it the
first-order tidal field in the LARF:
\begin{eqnarray} \label{eq:calcR}{\cal
E}^{(1)}_{ij} & \equiv & R^{(1)}_{0i0j}
 =  R_{0i0j} - R^{(0)}_{0i0j},
\end{eqnarray}
where a superscript of ``$(0)$'' or ``$(1)$'' indicates the quantity
is of zeroth or first order in the perturbation. In our
calculation, we found it convenient to look at the $0r0r$ component
of the first-order Riemann tensor in the LARF, since
\begin{equation}
R^{(1)}_{0r0r}
= R^{(1)}_{0i0j} n_i n_j
= {\cal E}^{(1)}_{ij} n_i n_j.
\end{equation}
From this equation we can read off ${\cal E}_{ij}^{(1)} =
R^{(1)}_{0i0j}$, the first-order tidal field in Cartesian
coordinates in the LARF, from the Riemann tensor in Schwarzschild
coordinates. By the procedure outlined in this paragraph we have
deduced the following $(\ell,m)=(2,2)$ part of ${\cal E}_{ij}^{(1)}$
in the LARF:
\begin{equation}
\label{eq:tide224}
{\cal E}^{(1),22}_{ij}
=
    -\frac{3\EuScript{A}}{4M^3}
    \left[1+\frac{4}{3}i\varepsilon +
        \sum_{n=5}^{\infty} \frac{i\varepsilon}{n\tilde{r}^n}
    \right] \mathscr{Y}^{22}_{ij} e^{-i\omega t}.
\end{equation}

The outer boundary condition states that the $O(r^0)$ [i.e.
$O(\tilde{r}^0)$] term of ${\cal E}^{(1),22}_{ij}$ must equal
$\mathcal{E}^{\text{ext},22}_{ij}$ [Eq.~(\ref{eq:tide22})]. This
determines the scaling factor to be
\begin{equation}
\label{eq:Aval}
\EuScript{A} =
    \frac{4 \mu M^3}{b^3}
    \sqrt{\frac{2 \pi}{15}}\left(1-\frac{4}{3}i\varepsilon\right).
\end{equation}
Inserting Eq.~(\ref{eq:Aval}) into Eq.~(\ref{eq:tide224}), we can
write $\mathcal{E}_{ij}^{(1),22}$ as
\begin{equation}
\label{eq:22res}
\mathcal{E}_{ij}^{(1),22} =
    {\cal E}^{\text{ext},22}_{ij} -
    \frac{\mu}{b^3}\sqrt{\frac{6\pi}{5}} \sum_{n=5}^{\infty}
    \frac{i\varepsilon}{n\tilde{r}^n}
    \mathscr{Y}^{22}_{ij}e^{-i\omega t}.
\end{equation}
Here the $O(1/\tilde{r}^5)$ term, by Eq.~(\ref{eq:tideGH}), contains
the induced quadrupole moment. The $O(1/\tilde{r}^6)$ and higher
terms are proportional to the $O(1/\tilde{r}^5)$ term and contain no
new information; they represent the non-linear coupling between the
induced quadrupole and the black hole's monopole moment.

Comparing the $O(1/\tilde{r}^5)$ term in Eq.~(\ref{eq:tideGH}) and
the $O(1/r^5)$ term in Eq.~(\ref{eq:22res}), we find that
\begin{equation}
\mathscr{I}^{22}_{ij} =
    \frac{32}{45}M^6 \mathcal{\dot{E}}_{ij}^{\text{ext},22}.
\end{equation}
Complex conjugating this equation yields the $(\ell,m)=(2,-2)$ part
of the induced quadrupole moment:
\begin{equation}
\mathscr{I}^{2\text{--}2}_{ij} =
    \frac{32}{45}M^6 \mathcal{\dot{E}}_{ij}^{\text{ext},2\text{--}2}.
\end{equation}
Thus, the time dependent part, i.e. the dynamical part (DP), of the
induced quadrupole moment is given by
\begin{equation}
\label{eq:resultForInducedMoment}
\mathscr{I}^{\text{ind, DP}}_{ij}
=   \frac{32}{45} M^6
    \Big(\mathcal{\dot{E}}_{ij}^{\text{ext},22} +
    \mathcal{\dot{E}}_{ij}^{\text{ext},2\text{--}2}\Big)
=
    \frac{32}{45} M^6 \mathcal{\dot{E}}_{ij}^{\text{ext}}.
\end{equation}
This agrees with the result recently obtained by
Poisson~\cite{poisson4} by a very different method. Note that the
induced quadrupole moment is proportional to the time derivative of
the applied tidal field. Hence the induced quadrupole moment and the
applied tidal field are completely out of phase with each other
($\pi/4$ phase shift in space, \(\pi/2\) in time). This leads to a
dissipative force acting back on the moon.

From the induced quadrupole
moment~(\ref{eq:resultForInducedMoment}), we define a corresponding
Newtonian potential in the LARF and out to the moon's orbit:
\begin{equation}\label{eq:pot}
\Phi= -\frac{3}{2}
       {\mathscr I}^{\text{ind, DP}}_{ij}
      \ \frac{n^i n^j}{r^3}.
\end{equation}
Then the force acting back on the moon can be found by evaluating
the gradient of $\Phi$ at the moon's position:
\begin{equation}
\label{eq:tangF}
{\mathbf F}
=
    -{\mathbf \nabla}\Phi|_{{\mathbf x}^\text{s}}
=
    -\frac{32}{5}\left(\frac{\mu}{b}\right)^{2}
    \left(\frac{M}{b}\right)^{13/2} {\mathbf e}_{\phi}.
\end{equation}
Equation~(\ref{eq:tangF}) shows that the force is tangential and
opposite to the moon's motion. The energy loss from the moon's
orbital motion is then
\begin{equation}
\label{eq:Eloss} {\dot E}
=   -\mathbf{F}\cdot \mathbf{v} =
    \frac{32}{5} M^4 \mu^2 \Omega^6,
\end{equation}
where $\mathbf{v}=\Omega b{\mathbf e}_\phi$. It is straightforward
to show that there is also an angular momentum loss of magnitude
$\dot{E}/\Omega$. Equation~(\ref{eq:Eloss}) agrees with Poisson and
Sasaki's calculation of the rate at which the perturbation carries
energy into the black hole's horizon at the leading post-Newtonian
order~\cite{poisson3}.

\section{The static, axisymmetric part of the perturbation}
\label{sec:static}

\subsection{Static induced quadrupole moment}
\label{sec:rwStatic} We now specialize to the even-parity, static
part of the moon's perturbation: $(\ell,m)=(2,0)$. The Regge-Wheeler
metric for this type of perturbation has the form~\cite{regge1}
\begin{eqnarray}
h^{(20)}_{ab}
 =
    \text{Diag}\Big[\Big(1-\frac{2M}{r}\Big){\hat H}_2,\
     \frac{{\hat H}_2}{1-2M/r}\ ,\qquad\qquad\quad \\
     r^2\hat{K}_2,\ r^2\hat{K}_2\sin^2\theta\Big]
    \times Y^{20}(\theta,\phi)
    \nonumber
\label{eq:metrich20}
\end{eqnarray}
where ``Diag" is short for diagonal matrix and ${\hat H}_2$ and ${\hat
K}_2$ are functions of $r$ only. The general solution to the field
equation governing ${\hat H}_2$ can be expressed in terms of the
associated Legendre functions~\cite{hartle1}:
\begin{equation}
\label{eq:hk20}
{\hat H}_2(r)
=   \alpha_2 P_{2}^{2}\left(\frac{r}{M}-1\right)
   +\beta_2  Q_{2}^{2}\left(\frac{r}{M}-1\right),
\end{equation}
where $\alpha_2$ and $\beta_2$ are constants to be determined. The
solution to $\hat{K}_2$ can then be obtained from that of
$\hat{H}_2$ (Appendix C). As $r$ approaches $2M$, we
have~\cite{transBook}
\begin{equation}
\label{eq:Q2m}
Q^{2}_{\ell}(r/M-1) \sim (r/M)^{-1/2}, \nonumber
\end{equation}
so the $Q^2_2$ term in Eq.~(\ref{eq:hk20}) becomes divergent at
$r=2M$ and we must set the coefficient $\beta_2$ to be zero in order
for the perturbation to be finite there. As $r$ goes to
infinity\footnote{Valid for all $\text{Re}\ \ell > -1/2.$},
\begin{equation}
\label{eq:Pinfty}
P^2_\ell(r/M-1) \sim (r/M)^{\ell}. \nonumber
\end{equation}
Therefore the remaining $P^2_2$ term in Eq.~(\ref{eq:hk20}) keeps
growing quadratically as $r$ becomes large, corresponding
to the non-asymptotic-flatness due to the presence of the moon.

With the metric perturbation $h^{(20)}_{ab}$, we compute the Riemann
tensor from the full metric and series expand the result up to
linear order in $\alpha_2$ (i.e., first order in the perturbation).
The $0r0r$ component of the resulting first-order Riemann tensor is
found to be
\begin{equation}
R^{(1)}_{0r0r}=\frac{3\alpha_2}{M^2}\ Y^{20}(\theta,\phi).
\end{equation}
From this and from Eq.~(\ref{eq:tide2m}), we obtain the first-order
tidal field in the Cartesian basis
\begin{equation}
{\cal E}^{(1)}_{ij}
=   \frac{3\alpha_2}{M^2}\ \mathscr{Y}_{ij}^{20}.
\label{eq:staticE}
\end{equation}
The static, first-order tidal field thus contains {\em only} an
$O(r^0)$ term, which should be identified as the static part of the
applied external field $\mathcal{E}_{ij}^{\text{ext},20}$
[Eq.~(\ref{eq:staticExtTide})]. The coefficient $\alpha_2$ is
determined from this identification to be $\alpha_2=\sqrt{4\pi/45}\
\mu M^2/b^3$. Since there is no $O(1/r^5)$ term present in
Eq.~(\ref{eq:staticE}), we infer that there is no static induced
quadrupole moment:
\begin{equation}
{\mathscr I}^{20}_{ij} = 0.
\label{eq:staticI}
\end{equation}

This is quite a counter-intuitive result. It is worth pointing out,
however, that the absence of negative powers of $r$ in
Eq.~(\ref{eq:staticE}) follows directly from the regularity
condition we imposed at $r=2M$. If the radius $r=2M$ were well
inside the central body itself, which naturally is the case for any
nonrelativistic body with weak self-gravity, then the $Q^2_2$ term
in Eq.~(\ref{eq:hk20}) would survive and give rise to an induced
quadrupole moment. Equation~(\ref{eq:staticI}) may also be the
consequence of the gauge (Regge-Wheeler) we choose to work in. Is it
possible to give a gauge-invariant definition of static, induced
multipole moment in a non-asymptotically-flat spacetime? This
is the question we shall investigate in the next subsection.

We now summarize and conclude that the total induced quadrupole moment
\emph{in our chosen gauge} is
\begin{equation}
{\mathscr I}^{\text{ind}}_{ij}
=   {\mathscr I}^{20}_{ij} + {\mathscr I}^{22}_{ij} +
    {\mathscr I}^{2\text{--}2}_{ij}
=   \frac{32}{45}M^6 \dot{\cal E}^{\text{ext}}_{ij},
\end{equation}
which is proportional to the time derivative of the external tidal
field --- not the field itself as one would expect for Newtonian
tidal couplings.

Lastly, we move from the LARF to the perturbed horizon and examine
the effect of the static perturbation there. Hartle has
shown~\cite{hartle1} that to first order in the perturbation, the
coordinate location of the event horizon of a slowly rotating black
hole perturbed by a stationary distribution of matter is still at
$r=2M$. This is also true for a Schwarzschild black hole under
static perturbations. Evaluating the full metric at $r=2M$, we find
the horizon metric is given by
\begin{equation}
\label{eq:staticHorizonMetric}
ds^2_{H} =
    4 M^2 \Big[1- 2\mu (M^2/b^3) P_2(\cos\theta)\Big]
    \left(d\theta^2 + \sin^2\theta d\phi^2
\right),
\end{equation}
where $P_2$ is the Legendre function. From this metric the scalar
curvature of the horizon is obtained as
\begin{equation}
\mathscr{R}=
    \frac{1}{2M^2}\Big[1 - 4\mu (M^2/b^3) P_2(\cos\theta)\Big].
\end{equation}
So it is clear that the shape of the horizon does acquire a small
quadrupolar component. But this deformation is {\em not} accompanied
by an induced quadrupole moment in the LARF, at least in our chosen
gauge.

\subsection{Ambiguity of the static induced quadrupole moment}
\label{sec:gauge}

In the previous subsection, we found that a Schwarzschild black hole
has a vanishing static induced quadrupole moment (SIQM) in response
to the external tidal field ${\cal E}^{\text{ext},20}_{ij}$.  To see
that this vanishing of the SIQM might possibly be a gauge effect,
imagine replacing the radial coordinate $r$ in the expression $\Phi
= (1/2) {\mathcal E}_{ij}^{\text{ext},20} n^i n^j r^2$ for the
external tidal Newtonian potential by $r= \bar{r} (1+ \chi
M^5/\bar{r}^5)^{1/2}$, where $\chi$ is some dimensionless number of
order unity.  The result is $\Phi = (1/2) {\mathcal
E}_{ij}^{\text{ext},20} n^i n^j \bar r^2 + (\chi/2) M^5 {\mathcal
E}_{ij}^{\text{ext},20} n^i n^j / \bar r^3$.  By comparing this
expression with Eq.\ (\ref{eq:tideGH}) we read off a SIQM
$\mathscr{I}_{ij} = (\chi/18) M^5 {\mathcal
E}_{ij}^{\text{ext},20}$. In Newtonian theory this procedure would
obviously be naive, but in general relativity, where the unperturbed
black hole metric can be expanded in powers of $M/r$ and the
coefficients in that expansion depend on one's choice of radial
coordinate and that choice is a ``gauge'' issue, this type of
procedure is not obviously naive at all.

From our point of view, the best way to explore the gauge dependence
of the SIQM is to ask whether it is physically measurable.  If (as
we shall find) physical measurements give a result that is ambiguous
at some level, then that ambiguity constitutes a sort of gauge
dependence of the SIQM.

In this section we shall study a thought experiment for measuring
the SIQM, one based on coupling to a small, static external ``test''
octupole field $\mathcal E_{ijk}$ (proportional to the symmetrized
and trace-removed gradient of some fiducial external quadrupolar
tidal field). For simplicity we take ${\mathcal E}_{ijk}$ to be
axisymmetric around the same $z$ axis as our static external tidal
field ${\cal E}^{\text{ext},20}_{ij}$; i.e.\ we take it to be
proportional to a tensor spherical harmonic of order $(\ell, m) =
(3,0)$:
\begin{equation}
{\cal E}_{ijk} \sim \mathscr{Y}^{30}_{ijk}. \nonumber
\end{equation}

The analysis in Ref.~\cite{zhang1985} says that, any SIQM
$\mathscr{I}_{ij}$ (created in the black hole by ${\cal
E}^{\text{ext},20}_{ij}$) will couple to the external octupole
moment to produce a force that gradually changes the hole's
momentum\footnote{The test octupole field may also induce a static
octupole moment \(\mathscr{I}_{ijk}\) in the central black
hole, which will couple to the external quadrupolar tidal field.
This coupling, as we shall show, contributes to the ambiguities in
the definition of the SIQM.}:
\begin{equation}
\dot{P}^i = -\frac{1}{2}\ {\cal E}^i_{\ jk} \mathscr{I}_{jk}.
\label{eq:EjabIab}
\end{equation}
(Eq.~(1.12) of Ref.~\cite{thorne3}; Eq.~(4b) of
Ref.~\cite{zhang1985}). The same will be true if the central black
hole is replaced by a neutron star or any other spherical body. The
rate of change of momentum $\dot{P}^i$ can also be evaluated by a
surface integral of the Landau-Lifshitz pseudotensor
$t^{ij}_\text{LL}$ in the LARF~\cite{thorne3}:
\begin{equation}
\dot{P}^i = -\oint (-g)\ t^{ij}_\text{LL}\ d S_j.
\label{eq:tLLds}
\end{equation}

Equations~(\ref{eq:EjabIab}) and (\ref{eq:tLLds}) for the
coupling-induced force on the hole actually have ambiguities that
arise from nonlinearities in the Einstein field equations. The
origin of those ambiguities is discussed with care in Sec.~I of
Thorne and Hartle \cite{thorne3}.  In this subsection we use
Eq.~(\ref{eq:tLLds}) to calculate the force on the hole, and shall
identify the ambiguities as those terms in which the force depends
on the location of the integration surface. The result of our
calculation will tell us, by comparison with Eq.~(\ref{eq:EjabIab}),
the SIQM and the amount by which it is ambiguous.

To compute the pseudotensor for insertion into Eq.~(\ref{eq:tLLds}),
we must solve for the metric perturbation containing both the
quadrupole and octupole terms:
\begin{eqnarray}
h_{ab}
&=& h^{(20)}_{ab} + h^{(30)}_{ab} \nonumber \\
&=&
    \sum\limits_{\ell=2,3} \text{Diag}\Big[
    \left(1-{2M/r}\right){\hat H}_\ell,\
     \frac{{\hat H}_\ell}{1-2M/r}\ , \\
& &
    \qquad\qquad\qquad
    r^2\hat{K}_\ell,\ r^2\hat{K}_\ell\sin^2\theta\Big]
    \times Y^{\ell 0}(\theta,\phi). \nonumber
\end{eqnarray}
When $\ell=2$, the general solution to $\hat{H}_2$ is given in
Eq.~(\ref{eq:hk20}). For $\ell=3$, we have
\begin{equation}
\label{eq:hk30}
{\hat H}_3(r)
=   \alpha_3 P_{3}^{2}\left(\frac{r}{M}-1\right)
   +\beta_3  Q_{3}^{2}\left(\frac{r}{M}-1\right).
\end{equation}
In order that both types of perturbation be finite at $r=2M$,
$\beta_2$ and $\beta_3$ must be set to zero [see
Eq.~(\ref{eq:Q2m})]. In order to deal with more general cases,
however, we keep {\em non-vanishing} values for $\beta_2$ and
$\beta_3$ in Eqs.~(\ref{eq:hk20}) and (\ref{eq:hk30}) so the
following analysis will be valid for stars as well as black holes.
[For central bodies other than black holes, $\beta_2$ and $\alpha_2$
(and similarly $\beta_3$ and $\alpha_3$) are not independent of each
other: $\beta_2$ is proportional to $\alpha_2$ with a
proportionality constant that depends on the body's internal
physical properties]. Having specified the metric perturbation, we
then insert the full metric into the expression for the pseudotensor
(Eq. (20.22) of Ref.~\cite{MTW})
\begin{eqnarray}
(-g)t^{\alpha\beta}_\text{LL}
&=&
    \frac{1}{16\pi}
    \Big\{
        \mathfrak{g}^{\alpha\beta}_{\ \ ,\lambda}
        \mathfrak{g}^{\lambda\mu}_{\ \ ,\mu}
    -
        \mathfrak{g}^{\alpha\lambda}_{\ \ ,\lambda}
        \mathfrak{g}^{\beta\mu}_{\ \ ,\mu}
    +
        g_{\lambda\mu}g^{\nu\rho}
        \mathfrak{g}^{\alpha\lambda}_{\ \ ,\nu}
        \mathfrak{g}^{\beta\mu}_{\ \ ,\rho}
    \nonumber \\
&-&
    \left(
        g^{\alpha\lambda}g_{\mu\nu}
        \mathfrak{g}^{\beta\nu}_{\ \ ,\rho}
        \mathfrak{g}^{\mu\rho}_{\ \ ,\lambda}
    +
        g^{\beta\lambda} g_{\mu\nu}
        \mathfrak{g}^{\alpha\nu}_{\ \ ,\rho}
        \mathfrak{g}^{\mu\rho}_{\ \ ,\lambda}
    \right) \nonumber \\
&+&
        \frac{1}{2}g^{\alpha\beta} g_{\lambda\mu}
        \mathfrak{g}^{\lambda\nu}_{\ \ ,\rho}
        \mathfrak{g}^{\rho\mu}_{\ \ ,\nu}
    +
        \frac{1}{8}\left(
            2g^{\alpha\lambda}g^{\beta\mu} -
            g^{\alpha\beta}g^{\lambda\mu}
        \right)
    \nonumber\\
&\times&
        \left(
            2g_{\nu\rho}g_{\sigma\tau} -
            g_{\rho\sigma} g_{\nu\tau}
        \right)
        \mathfrak{g}^{\nu\tau}_{\ \ ,\lambda}
        \mathfrak{g}^{\rho\sigma}_{\ \ ,\mu}
    \Big\},
\end{eqnarray}
and evaluate the surface integral at some radius $r=R$ in the LARF.
Because of the axi-symmetry of the perturbed spacetime, only the
$z$-component of $\dot{P}^i$ is nonzero. The result, up to
first-order coupling and with uninteresting numerical coefficients
being suppressed, has the following form:
\begin{eqnarray}
\label{eq:pdotz}
\dot{P}^z
&=&
    \alpha_3\alpha_2\left[
        \frac{R^4}{M^4}\ \&\ \frac{R^3}{M^3}\ \&\ \frac{R^2}{M^2}
        \ \&\ \frac{R}{M}\ \&\ 1\ \&\ ... \right] \\
&+&
    \alpha_3\beta_2\left[1\ \&\ \frac{M}{R}\ \&\ ... \right]
    +\beta_3\alpha_2
     \left[\frac{M^3}{R^3}\ \&\ \frac{M^4}{R^4}\ \&\ ...\right],
     \nonumber
\end{eqnarray}
where ``\&" is to be read ``and a term of the order".

The constant terms in Eq.~(\ref{eq:pdotz}) [i.e., the ``1''s] that
are independent of the integration radius $R$ are the ones to be
compared with Zhang's result~(\ref{eq:EjabIab}) so as to deduced the
gauge-invariant SIQM. Other terms that depend on $R$ constitute
ambiguities\footnote{The $M^2/R^2$ term includes the effect of any
octupole moment induced on the central body by the test octupole
field. Note that $\dot{P}^i$ is a dimensionless vector. On
dimensional grounds, the coupling between any induced octupole
moment and the external tidal field must take the form
$\mathscr{I}^{ijk}{\cal E}_{jk}/R^2$ to contribute to $\dot{P}^i$.
(Nonlinear coupling to the monopole moment can lead to similar terms
that scale as higher, but not lower, powers of \(1/R\)).}. Terms
with positive power(s) of $R/M$ appear because the spacetime is not
asymptotically flat, and they prevent us from minimizing the
ambiguities by simply pushing the integration surface to infinity.

Let us step back and write down the most general form that the SIQM
can take. By order of magnitude analysis of the response of any
physical body (black hole, star, planet, moon, ...) to a tidal
field, we must have
\begin{eqnarray}\label{eq:oom1}
\mathscr{I}_{ij} \sim
    L^5 (1 + \xi)\ {\cal E}^{\text{ext},20}_{ij} .
\end{eqnarray}
Here $L$ is the size of the body ($L\sim M$ for a black hole) and
$\xi$ is a dimensionless number describing the SIQM's dependence on
the integration radius $R$ --- deviations from being well-defined.
From Eq.~(\ref{eq:staticE}), we know the external tidal field scales
as $\sim\alpha_2/M^2$. Similarly for the external octupole field,
${\cal E}_{ijk}\sim\alpha_3/M^3$. Using these relations,
Eq.~(\ref{eq:EjabIab}) becomes
\begin{eqnarray}
\dot{P}^i
&\sim&
    L^5 (1 + \xi)
    \ {\cal E}^{\text{ext},20}_{jk}\ {\cal E}^{i}_{\ jk}\nonumber\\
&\sim&
    \alpha_2\alpha_3
    \left[ \frac{L^5}{M^5} + \frac{\xi L^5}{M^5}\right].
\end{eqnarray}
Here the first term in the square bracket should be identified
as the ``1''s in Eq.~(\ref{eq:pdotz}) [note again that $\beta_2$ and
$\alpha_2$ are not independent of each other for stars]; and the
second term should be identified as the sum of all $R$-dependent terms:
\begin{equation}
\label{eq:xiSum}
\frac{\xi L^5}{M^5} =
    \frac{R^4}{M^4}\ \&\ ...\ \&\ \frac{R}{M}\
    \&\ \frac{M}{R}\ \&\ ... .
\end{equation}

In the case of a black hole we have $L\sim M$ and the smallest the
right hand side of Eq.~(\ref{eq:xiSum}) can be is $\sim 1$ (for
$R\sim M$), so $\xi \gtrsim 1$, i.e. \emph{the SIQM for a
Schwarzschild black hole is ambiguous by an amount} $\gtrsim M^5
\mathcal{E}_{ij}^{\text{ext},20}$, i.e. \emph{totally ambiguous},
since the largest we could expect $\mathscr{I}_{ij}$ to be is $\sim
M^5 \mathcal{E}_{ij}^{\text{ext},20}$.

For central objects with $L \gg M$ (e.g., the Earth) we must choose
$R > L$. The right hand side of Eq.~(\ref{eq:xiSum}) is then minimized
by setting $R \simeq L$, giving $\xi \sim M/L \ll 1$ $(\sim 10^{-9}$
in the case of the Earth) for the fractional ambiguities in the
SIQM.

We comment that our result for a Schwarzschild black hole differs
from what Suen has derived. Suen has given an unambiguous
prescription to read out static multipole moments in
non-asymptotically-flat spacetimes, which is based on transforming
coordinates into a particular set of de Donder
coordinates~\cite{suen1}. He has used his prescription to calculate
the induced quadrupole moment of a Schwarzschild black hole when it
is perturbed by a static, equatorial matter ring at large
distances~\cite{suen2}. According to his prescription, the SIQM does
{\em not} vanish. It is proportional to the tidal field produced by
the ring:
\begin{equation}
\label{eq:suen} {\mathscr I}_{ij} =-\frac{4}{21} M^5
{\cal E}^{\text{ring}}_{ij}.
\end{equation}
The incompatibility between this result and the vanishing SIQM that
we derived in Sec.~\ref{sec:rwStatic} in Regge-Wheeler gauge and
Schwarzschild coordinates illustrates the ambiguities of the SIQM.
Both results, zero and $(-4/21) M^5 \mathcal{E}_{ij}$ are less than
or of order the ambiguity.

\section{The tidal phase shift}
\label{sec:phase} In the LARF, the time-dependent induced quadrupole
moment is \(\pi/4\) out of phase with the perturbing tidal field
(Sec.~\ref{sec:tdmoment}). This large phase shift is quite different
from the small phase lag angle, caused by viscous dissipation,
between a planet's induced quadrupole moment and the perturbing
tidal field. A closer black-hole analogy to a planet's viscous phase
lag may be found by considering the tide raised on the hole's
horizon by an orbiting moon.

In this section, we compute the tidal phase shift on the horizon for
our perturbed Schwarzschild black hole. We will discuss in what
sense it is and is not analogous to the fluid-planet's viscous phase
lag. To calculate this phase shift, it is convenient to use the
Newman-Penrose formalism~\cite{np} (see, e.g., chapter 1 of
Ref.~\cite{chandraBook} for a review of the Newman-Penrose
formalism). Appendix~\ref{sec:npAp} summarizes some details of the
Newman-Penrose formalism that are relevant for our purpose.

We consider two approaches to defining the tidal phase shift. In
Sec.~\ref{sec:mpphase}, we define the phase shift only in terms of
quantities on the horizon (following the method suggested in
Sec.~VIIC of Ref.~\cite{MPBook}), while in
Sec.~\ref{sec:hartlephase}, we define a phase between the tide
raised on the horizon and the ``retarded'' position of the moon
(following the method used by Hartle in Ref.~\cite{hartle2}).

\subsection{Phase of the tidal bulge on the horizon}
\label{sec:mpphase}

\emph{For Sec.~\ref{sec:phase} and Appendix~\ref{sec:npAp} only}, we
use ingoing Eddington-Finkelstein coordinates
\((\tilde{V},r,\theta,\phi)\) and a \((+~-~-~-)\) signature of the
metric. The Schwarzschild metric in these coordinates is
\begin{equation}\label{eq:SchwarzschildInEF}
ds^2
= \left(1-\frac{2 M}{r}\right) d\tilde{V}^2 - 2 d\tilde{V} dr -
  r^2 \left(d\theta^2 + \sin^2 \theta d\phi^2\right).
\end{equation}
The ingoing Eddington-Finkelstein null time coordinate $\tilde V$ is
related to the Schwarzschild time coordinate $t$ and radial
coordinate $r$ by the following equation (Eq.~(1b) of Box 32.2 of
Ref.~\cite{MTW}):
\begin{equation}
\tilde V = t + r^* = t + r + 2M \ln |r/2M -1|\;,
\label{eq:tildeV}
\end{equation}
and the Eddington-Finkelstein radial and angular coordinates
$\{r,\theta,\phi\}$ are identical to those of Schwarzschild.

Our slowly orbiting moon deforms the Schwarzschild event horizon. By
analogy with Newtonian tides, we would like to describe the horizon
deformation as a perturbation that co-rotates (at a slightly
different phase) with the tidal field that drives it. But this
viewpoint inherently envisions the perturbed event horizon as a
two-dimensional, evolving surface, rather than as a
three-dimensional, global surface in spacetime. Therefore, before we
can consider the \emph{phases} of quantities on the horizon, we must
first specify what we mean by \emph{time} on the horizon.

Begin by considering the Schwarzschild event horizon (which is, of
course, the three-surface \(r=2M\)). There is a preferred way to
slice the horizon into a single-parameter family of two-surfaces;
this preferred slicing uses two-surfaces that are orthogonal to the
Schwarzschild Killing vector \(\partial_t = \partial_{\tilde{V}}\)
that is timelike at the moon and null on the horizon. Following
Hartle~\cite{hartle2}, we call this family of two-surfaces the
``instantaneous horizon''. The instantaneous horizon can be pictured
as an evolving two-surface defined by \(r=2M\) and
\(\tilde{V}=\text{const}\), so that \(\tilde{V}\) plays the role of
a ``time'' coordinate. Throughout this section, we use the terms
``horizon'' and ``instantaneous horizon'' interchangeably unless
otherwise indicated.

We now consider how the horizon's perturbation evolves with time
\(\tilde{V}\). The moon's tidal field, characterized in the LARF by
[Eq.~(\ref{eq:tide2m})]
\begin{equation}
\mathcal{E}_{ij}^{\text{ext}}
=
    \mathcal{E}_{ij}^{\text{ext},20} +
    \mathcal{E}^{\text{ext},22}_{ij}+
    \mathcal{E}_{ij}^{\text{ext},\text{2--2}}, \nonumber
\end{equation}
deforms the otherwise spherical, static horizon. Because
\(\mathcal{E}_{ij}^{\text{ext},20}\) is static and axisymmetric, it
cannot contribute to the phase shift. For the remaining tidal
fields, \(\mathcal{E}_{ij}^{\text{ext},2\pm2}\), we shall consider
only the \((2,2)\) mode in detail and the result for the \((2,-2)\)
mode follows immediately.

On the horizon, it is the tangential-tangential components of the
perturbing tidal field that drive the deformation (see, e.g.,
Eq.~(6.80) of Ref.~\cite{MPBook}); knowledge of these components is
physically equivalent to knowledge of the Teukolsky function
\(\Psi_0\) \cite{Teukolsky72} (see, e.g., Eq.~(A7) of
Ref.\cite{price1}). The Teukolsky function is a particular component
of the Weyl tensor [Eq.~(\ref{eq:psi0def})].

The horizon deformation is governed by the Newman-Penrose equation
(Eq.~(2.11) of Ref.~\cite{hartle2})
\begin{eqnarray}\label{eq:tidalForceEq}
(\partial_{\tilde{V}}-2\epsilon)\sigma^{(1)}
= -2(i\Omega+\epsilon)\sigma^{(1)} = \Psi_0^{(1)}.
\end{eqnarray}
This is also the ``tidal force equation'' (equation~(6.80) of
Ref.~\cite{MPBook}). Here \(\sigma=-\Sigma^{(1)}\) is a
Newman-Penrose spin coefficient [Eq.~(\ref{eq:spinDefs})] and
\(\Sigma^{(1)}\) is the shear (i.e., the rate of change of the
shape\footnote{Recall that the shape of the instantaneous horizon (a
two-dimensional surface) is completely specified by its intrinsic
scalar curvature \(\mathscr{R}\).}) of the instantaneous horizon.
Note that because \(\Sigma\) and \(\Psi_0\) vanish on the
unperturbed instantaneous horizon, the spin coefficient \(\epsilon\)
takes its Schwarzschild value, which (in our tetrad) is the surface
gravity of the instantaneous horizon \(g_H = (4 M)^{-1}\).

Knowing \(\Psi_0^{(1)}\), we can evaluate the horizon shear. Because
\(\Psi_0^{(1)}\) is first order in the perturbation, it may be
evaluated on the horizon simply by letting $r$ go to $2M$.

Beginning with the \((\ell,m)=(2,2)\) metric perturbation
[Eq.~(\ref{eq:mp22}), except we now choose the  metric signature to
be \((+~-~-~-)\)], we compute the perturbed Riemann tensor near the
horizon and read off the component \(\Psi_0^{(1)}\). The result is
\begin{eqnarray}\label{eq:psi0}
\Psi_0^{(1)} & = & -i\sqrt{\frac{\pi}{5}} \frac{\mu M
\Omega}{b^3} ~_{2}Y^{22} e^{-2 i \Omega \tilde{V}+(8/3)M\Omega} + O(M^2\Omega^2)\nonumber\\ & = &
|\Psi_0^{(1)} | \exp\left[2 i \left(\phi - \Omega \tilde{V} + \frac{4}{3}M\Omega - \frac{\pi}{4} \right)\right] \nonumber\\ & & + O(M^2\Omega^2).
\end{eqnarray}
Here \(~_2 Y^{2 2}\) is the spin-weighted spherical harmonic
\begin{eqnarray}
~_2 Y^{22}
=   \frac{1}{2} \sqrt{\frac{5}{\pi}} \sin^4\left(
    \frac{\theta}{2}\right) e^{2 i \phi}. \nonumber
\end{eqnarray}

With \(\Psi_0^{(1)}\) in hand, we can calculate \(\Sigma^{(1)}\) via
Eq.~(\ref{eq:tidalForceEq}). Inserting \(\epsilon = 2 g_H\) and
\(\omega = 2\Omega\) into Eq.~(\ref{eq:tidalForceEq}) yields
\begin{eqnarray}
\label{eq:shear1}
\Sigma^{(1)}
 & = &  \frac{\Psi_0^{(1)}}{i\omega+2\epsilon}\nonumber\\
&=&
    4M\Psi_0^{(1)}e^{-2 i \Omega/g_H } + O(M^2 \Omega^2)\nonumber\\
& = &
    |\Sigma^{(1)}| \exp\left[2 i \left(\phi - \Omega \tilde{V} +\frac{4}{3} M\Omega - \frac{\pi}{4} - \delta_\text{Horizon}\right)\right] \nonumber\\ & & + O(M^2\Omega^2).
\end{eqnarray}
where
\begin{equation}
\label{eq:deltaHor}
\delta_\text{Horizon} \equiv \Omega/g_H = 4M\Omega.
\end{equation}
\emph{The shear} \(\Sigma^{(1)}\) \emph{leads} \(\Psi_0^{(1)}\) (or
equivalently, the perturbing tidal field at the horizon) \emph{by an
angle} \(\delta_\text{Horizon}\). [Note that the first equality in
Eq.~(\ref{eq:shear1}) appears in Ref.~\cite{hartle2} as Eq.~(2.12).]

The shear is the time derivative of the shape. Therefore, the shape
has a phase
\begin{equation}
\label{eq:shapeExpect} \mathscr{R}^{(1)}
= |\mathscr{R}^{(1)}| \exp\left[ 2i\left(\phi-\Omega \tilde{V} - \frac{8}{3}M\Omega\right)\right].
\end{equation}
In other words, \emph{the shear leads the shape by} \(\pi/4\).

The horizon phase shift in Eq.~(\ref{eq:deltaHor}) follows directly
from the tidal force equation (\ref{eq:tidalForceEq}). It is
gauge-invariant since it only makes reference to gauge-invariant
quantities measured on the instantaneous horizon. In
Ref.~\cite{MPBook} [Eq.~(7.45), Fig.~57, and the surrounding discussion], an analogous horizon phase shift \(\delta_{\text{Horizon}}^\text{H}\) was deduced from the tidal force
equation for a slowly rotating black hole perturbed by a stationary,
axisymmetric tidal field---physically the same problem as Hartle
studied ~\cite{hartle2}:
\begin{equation}
\delta^\text{H}_{\text{Horizon}} = \Omega_H/g_H=4M\Omega_H = \delta_{\text{Horzion}}\left. \right|_{\Omega\rightarrow\Omega_H}.
\end{equation} Here $\Omega_H$ is the horizon angular velocity. 

Although Hartle also used the tidal force equation in his
calculations, he chose to define the tidal phase shift in a
different way and made his result gauge-invariant by making a
connection between the angular positions on the horizon and angular
positions at infinity through a null ray---a choice we will consider
in detail in Sec. \ref{sec:hartlesummary} and apply to our problem in Sec. \ref{sec:useHartle}.

The phase lead \(\delta_\text{Horizon}\) is, in some ways, analogous
to the phase shift of a tide raised on a non-rotating fluid planet.
In the latter case, viscous dissipation causes the shape of the
planet's surface to lag the normal-normal component of the
perturbing tidal field by a small angle \(\delta_\text{visc}\);
somewhat analogously, the horizon shear leads the
tangential-tangential component of the perturbing tidal field. Both
phase shifts are small angles associated with dissipation (which
manifests itself as a secular evolution of the energy and angular
momentum of the moon's orbit). In the absence of dissipation, there
is no phase shift. On the other hand, the phase shift
\(\delta_\text{Horizon}\) is a \emph{lead} angle while $\delta_\text{visc}$
is a \emph{lag} angle. Hartle explains this difference as a consequence of
the teleological nature of the horizon~\cite{hartle2}. Also as
Hartle observed, when the angular velocity \(\Omega\) is not small
compared with $1/M$, the deformation of the horizon cannot be
described in terms of a phase shift~\cite{hartle2}.

\subsection{Phase shift between the tidal bulge and the moon}
\label{sec:hartlephase}

As an alternative to the above way of defining the tidal phase
shift, one can define it as the angle between the tidal bulge on
the horizon and the location of the moon in its orbit. Hartle used
this approach when he computed the tidal lead on a rotating hole
perturbed by a stationary moon~\cite{hartle2}. First, we will briefly summarize the aspects of Hartle's analysis which are relevant to our purpose. Then, we will apply his method to a slowly rotating moon around an otherwise Schwarzschild black hole.

\subsubsection{Tidal phase shift between a rotating horizon and stationary moon}
\label{sec:hartlesummary}
In Ref. \cite{hartle2}, Hartle considers the problem of a distant, stationary moon perturbing a slowly rotating black hole. 


The Kerr metric can be written as
\begin{eqnarray}\label{eq:mKerrInEF}\label{eq:mKerrHH}
ds^2 & = & \left(1-\frac{2 M r}{\Sigma}\right)d\tilde{V}^2 - 2 d\tilde{V} dr + \frac{4 a M r \sin^2\theta}{\Sigma} d\tilde{V}d\tilde{\phi}\nonumber\\
& & + 2 a \sin^2\theta dr d\tilde{\phi} - \Sigma d\theta^2  \nonumber\\
& & - \sin^2\theta \left( a^2 + r^2 + \frac{2 a^2 M r \sin^2\theta}{\Sigma} \right) d\tilde{\phi}^2
\end{eqnarray} Here \(\Sigma \equiv r^2 + a^2 \cos^2\theta\). The coordinates \(\tilde{V}\) and \(\tilde{\phi}\) are related to the usual Boyer-Lindquist coordinates \(t\) and \(\phi\) by
\begin{eqnarray}
dt & = & d\tilde{V} - \frac{r^2+a^2}{\Delta} dr \nonumber\\
d\phi & = & d\tilde{\phi} - \frac{a}{\Delta} dr
\end{eqnarray} where \(\Delta \equiv r^2 - 2 M r + a^2\). When \(a=0\), Eq.~(\ref{eq:mKerrInEF}) reduces to the Schwarzschild metric in Eddington-Finkelstein coordinates [Eq.~(\ref{eq:SchwarzschildInEF})].

The event horizon is the surface \(r=r_+\equiv M+\sqrt{M^2-a^2}\). Just as in the Schwarzschild case considered above, the event horizon can be sliced into a single-parameter family of two-dimensional surfaces using the Killing vector \(\partial_{\tilde{V}}\) which is timelike at infinity and null on the horizon. This family of surfaces is the instantaneous horizon.

The distant moon raises a tidal bulge on the central hole's instantaneous horizon. In the limit that the moon is far away, the change in the horizon's shape (or equivalently, the change in the scalar curvature \(\mathscr{R}\) of the instantaneous horizon), is purely quadrupolar.

The deformation is driven by the transverse-transverse component of the tidal field at the horizon, which is physically equivalent to the Teukolsky function, a particular component of the Riemann tensor \(\Psi_0\) [Eq.~(\ref{eq:psi0def})]. This component vanishes in the unperturbed Kerr spacetime [Eq.~(\ref{eq:psiVanishKerr})], and the first order correction \(\Psi_0^{(1)}\) has the form
\begin{equation}
\Psi_0^{(1)} = S^{\ell m}(r) ~_2 Y^{\ell m}(\theta,\tilde{\phi})
\end{equation} where \( ~_2 Y^{\ell m}\) is a spin-weight-2 spherical harmonic. Because the perturbation is purely quadrupolar, we need only consider the case \(\ell=2,m=2\) here, although Hartle considers the generic case. Hartle uses Teukolsky's solution \cite{teukolskyThesis} for the stationary radial functions \(S^{\ell m}\) due to the \(\ell\)-pole perturbation caused by a distant, stationary point particle with mass \(\mu\). Furthermore, while Hartle treats the case of a moon at any location \((\theta,\tilde{\phi)}\), for concreteness we specify the moon's position as \((\theta,\tilde{\phi})=(\pi/2,0)\). On the horizon, the Teukolsky function turns out to have the value (combining Eqs.~(4.30), (4.31), (4.15), and (4.18) of Ref.~\cite{hartle2})
\begin{eqnarray}
\Psi_0^{(1),\text{H}} & = & 
\frac{i \mu M \Omega_H}{2 \sqrt{6} b^3}\sin^4\left(\frac{\theta}{2}\right)
\exp\left[2 i \left(\tilde{\phi}+2 M \Omega_H\right)\right] \nonumber\\ & & + O\left(\frac{M^4}{b^4}\right) + O\left(M^2\Omega_H^2\right).
\end{eqnarray}

The tidal field deforms the instantaneous horizon, changing it's shape and thus its two-dimensional scalar curvature \(\mathscr{R}\). Hartle computes the quadrupolar correction to the scalar curvature, \(\mathscr{R}^{(1),\ell=2,\text{H}}\), of the instantaneous horizon [Eq.~(\ref{eq:shapeFromTide})]. His result is (Eqs.~(4.26)--(4.27) of Ref.~\cite{hartle2})
\begin{eqnarray}
\mathscr{R}^{(1),\ell=2,\text{H}}  & \propto &  
\cos\left[2\left(\tilde{\phi}+\frac{14}{3} M \Omega_H\right)\right]\nonumber\\ & &
+ O\left(\frac{M^4}{b^4}\right) + O\left(M^2 \Omega_H^2\right).
\end{eqnarray}

Instead of measuring the angle between the shear \(\sigma\) and the tidal field \(\Psi_0\) on the horizon, Hartle defines his phase lead as the angle between the shape and the moon's angular position. To make this definition gauge-invariant, Hartle chooses ingoing, zero-angular-monentum, null geodesics to be ``lines of constant angle.'' He then compares the angular position of the horizon tidal bulge,
\begin{equation}
\tilde{\phi}_\text{bulge}^\text{H} = -\frac{14}{3} M \Omega_H.
\end{equation}
to the angular position of the moon on the horizon.

Consider stationary moon in the equatorial plane at (large) radius \(r=b\) and at angular position \(\phi=0\). An ingoing null ray, originating from the moon, intersects the instantaneous horizon at angular position\footnote{\label{sec:signError} Note that there is a sign error in Hartle's analysis. Hartle incorrectly states that the ingoing null ray intersects the horizon at \(+a/2M+O(a/b)\), not \(-a/2M+O(a/b)\). Had we also made this error, there would be a coefficient of \(20/3\) instead of \(8/3\) in Eq.~(\ref{eq:hartlePhaseRes}).
}
\begin{equation}
\tilde{\phi}_{\text{moon}}^{\text{H}} =a/b - a/2M. 
\end{equation} The tidal bulge therefore \emph{leads} the moon's position by an amount
\begin{eqnarray}\label{eq:hartlePhaseRes}
\delta_{\text{null map}}^\text{H} & = & \tilde{\phi}_\text{moon}^\text{H}-\tilde{\phi}_\text{bulge}^\text{H}\nonumber\\ & = & \frac{8}{3} M \Omega_H + 4\frac{M^2 \Omega_H}{b}.
\end{eqnarray} Here we have used the relation (valid for small \(a/M\)) that \(a = 4 M^2 \Omega_H\), with \(\Omega_H\) being the angular velocity of the hole. For simplicity, one can then take the limit \(b\rightarrow\infty\). 

Before continuing, we should remark that Hartle's prescription for constructing \(\delta_{\text{null map}}^{\text{H}}\) can be described without reference to the moon's position. Begin by computing the angular location of the tidal bulge on the horizon. Next, ingoing, zero-angular-momentum null rays from infinity define lines of constant angle, so that there is a one-to-one correspondence between angular positions on the horizon and angular positions at infinity. The angular position at infinity of the tidal bulge can thus be computed. Finally, perform the calculation again, but this time perturb a \emph{non-rotating spacetime}; in this case, there will be no tidal friction. Because the Kerr and Schwarzschild spacetimes are asymptotically identical, one can unambiguously compare the angular position of the tidal bulge in the presence and in the absence of tidal friction: \(\delta_{\text{Null Map}}^\text{H} = \phi^\text{H}_\text{bulge} - \phi^\text{H}_{\text{bulge, no friction}}\). This is equivalent to the previous definition of \(\delta_{\text{null map}}^\text{H}\) provided that \(b\rightarrow\infty\).

However, this alternative formulation of \(\delta_{\text{null map}}^\text{H}\) breaks down when the moon, not the horizon, rotates. The rotation is then described by \(\Omega\), which is a parameter of the perturbation, not of the background spacetime. To eliminate tidal friction, one must let \(\Omega\rightarrow 0\), which eliminates the perturbation\footnote{Even if \(\Omega\rightarrow 0\) resulted in a non-zero perturbation, it is unclear how to distinguish such a perturbation from a small change in the coordinates of the background spacetime.}. Because of this failure, we prefer to consider Hartle's phase shift as a comparison of the position of the tidal bulge with the position of the moon.  

\subsubsection{Tidal phase shift between a non-rotating horizon and rotating moon}
\label{sec:useHartle}
A similar analysis can be applied to our system, in which a distant moon in a slow, circular orbit raises a tide on a non-rotating black hole.

The moon orbits the central black hole along the world line
specified by Eq.~(\ref{eq:orbit}). In other words, the moon has a
phase given by
\begin{equation}
\label{eq:phimoon}
\phi_{\rm moon}(\tilde V) \equiv \Omega t
= \Omega (\tilde V - b^*).
\end{equation}
This must be compared with the location of the bulge on the hole's
future horizon. Equation~(\ref{eq:shapeExpect}) for
\(\mathscr{R}^{(1)}\) [or, alternatively, inserting Eq.~(\ref{eq:psi0}) into Eq.~(\ref{eq:shapeFromTide})] shows that the tip of the tidal bulge has a
phase given by
\begin{equation}\label{eq:phibulge}
\phi_{\rm bulge} = \Omega\tilde{V} + \frac{8}{3} M \Omega.
\end{equation}
As time $\tilde V$ passes, this bulge rotates around and around the
horizon, with the same angular velocity $\Omega$ as the moon that
raises the tide.

Following Hartle, we compare the angular location of the tidal
bulge, $\phi_{\rm bulge}(\tilde V)$, with the angular location of
the moon, $\phi_{\rm moon}(\tilde V)$, using ingoing, zero angular
momentum (ZAM) null rays to provide the connection between $\phi$ at
the moon's orbit and $\phi$ on the horizon.  In the ingoing
Eddington-Finkelstein coordinates that we are using, these ZAM rays
have a very simple form:
\begin{equation}
\{\tilde V, \theta, \phi\} =
\text{constant}\;, \ r \text{ decreases from $b$ to }2M\;.
\label{rays}
\end{equation}
Since $\tilde V$, $\theta$ and $\phi$ are all constant along these
rays, they give us a one-to-one map of events $\{\tilde V, r=b,
\theta, \phi\}$ at the moon's orbital radius to events $\{\tilde V,
r=2M, \theta, \phi\}$ on the horizon that have identically the same
$\tilde V$, $\theta$, and $\phi$.  With the aid of this map, we
conclude that the angle by which the horizon bulge lags the moon's
position is
\begin{eqnarray}
\delta_\text{null map}
\equiv
    \phi_{\rm bulge}(\tilde V) - \phi_{\rm moon}(\tilde{V})
 =
    \frac{8}{3} M \Omega + \Omega b^*.
\label{delta}
\end{eqnarray}
[Eqs.~(\ref{eq:phimoon}) and (\ref{eq:phibulge})]. Again, the phase
shift is a phase lead, not a phase lag, due to the teleological
nature of the horizon.

In addition to the teleological phase shift of order \(M\Omega\),
\(\delta_\text{null map}\) contains a much larger term of magnitude
\(\Omega b^*\); this term reflects the choice to use an
ingoing-null-ray mapping between the moon and the
horizon. A similar term appears in Hartle's
calculation [Eq.~(\ref{eq:hartlePhaseRes})], but in Hartle's system the term is much \emph{smaller} than the teleological
phase shift size (specifically, smaller by a factor of \(M/b\)), whereas \(\Omega b^* \gg M\Omega\).

One could avoid this problem by \emph{defining} the phase shift to include only terms of order \(M\Omega\) and \(M \Omega_H\). With this definition, the remaining tidal phase leads are the same: \((8/3) M \Omega\), as one would expect, given that a there should be no tidal shift at all if the moon were to rotate at the hole's angular velocity, i.e., if \(\Omega=\Omega_H\).

We prefer, however, to define the tidal lead angle in as the angle \(\delta_{\text{Horizon}}=4 M \Omega\) by which the horizon shear leads the horizon tidal field. This angle, in contrast to \(\delta_\text{null map}\), is defined in
terms of an ``instantaneous'' (spacelike) connection between the moon and the
horizon, i.e., by the near zone mapping of the moon's position to the horizon tidal field's [\(\Psi_0^{(1)}\)'s] maximum. [Had the moon been in the radiation zone (\(b \gg \lambda/2\pi\)), one would have expected the connection to be lightlike.]

\section{Conclusion}
\label{sec:conclusion} For our simple system of a Schwarzschild
black hole and circularly orbiting moon, we have found that the
time-dependent part of the moon's tidal field induces a quadrupole
moment that is unambiguous. The static induced quadrupole moment was
found to be zero in the Regge-Wheeler gauge, but it is ambiguous in
general. The ambiguity of the static induced quadrupole moment leads
to an ambiguity in the phase of the induced quadrupole moment in the
LARF; however, the tidal bulge on the horizon still has a well
defined phase shift with respect to the orbiting moon. Because of
the ambiguity of the induced quadrupole moment and the LARF phase
shift, we conclude that the polarizability and phase shift are not
suitable for constructing a body-independent description of tidal
coupling in EMRIs.

However, this conclusion does not eliminate the possibility of developing a body-independent language to describe tidal coupling, including cases where the central body is a black hole. It might be possible, for instance, to define a new set of induced ``dissipative multipole moments'' for the central body --- i.e. moments that vanish in the absence of tidal friction. Such dissipative moments would still be linear in the perturbing tidal field, so one could still define a polarizability. Also, by ignoring any non-dissipative tidal coupling, the phase shift might no longer contain additional information. Even if such an extension does not prove feasible, tidal coupling can still be described in the more conventional (but still body-independent) language of energy and angular momentum transfer between the moon and the central body. 

Other future work could
include generalizing our analysis to spinning black holes, treating noncircular, non-equatorial orbits, and (most importantly) studying how information about tidal coupling in EMRIs can be extracted from the gravitational waves detected by LISA.

\begin{acknowledgments}
We are grateful to Kip Thorne for suggesting this problem and for
his encouragement and advice, and to Yasushi Mino for helpful
discussions. This research was supported in part by NASA grants
NAG5-12834 and NAG5-10707 and NSF grant PNY-0099568. We used
\textit{Mathematica} and Maple to verify some of the equations in
this paper.
\end{acknowledgments}

\appendix

\section{Symmetric trace-free tensor notation for spherical harmonics}
\label{sec:ylmref} The scalar spherical harmonics $Y^{\ell
m}(\theta,\phi)$ can be written in terms of of rank-$\ell$ symmetric
trace-free (STF) tensors~\cite{thorne1}. The spherical harmonics
$Y^{2m}(\theta,\phi)$ that have been used in this paper are
\begin{subequations}
\begin{eqnarray}
Y^{2\pm 2}(\theta,\phi) =
\frac{1}{4}\sqrt{\frac{15}{2\pi}}\sin^{2}\theta e^{\pm 2 i
\phi}\label{eq:Y22s}\\
Y^{20}(\theta,\phi) =
\frac{1}{8}\sqrt{\frac{5}{\pi}}\left(1+3\cos
2\theta\right)\label{eq:Y20s}.
\end{eqnarray}
\end{subequations}
They can be written in terms of rank-2 STF tensors as (Eq.~(2.11) of
Ref.~\cite{thorne1})
\begin{equation}\label{eq:stfconvert}
Y^{2 m}(\theta,\phi) = \mathscr{Y}^{2 m}_{ij} n^{i} n^{j},
\end{equation}
where $n^{i} \equiv x^{i}/r$ and $\mathscr{Y}^{2 m}_{ij}$ are the
STF tensors given by (Eq.~(2.12) of Ref.~\cite{thorne1}):
\begin{eqnarray}
{\mathscr Y}^{20}_{ij}\label{eq:y20tensor}
    &=&
       -\frac{1}{4}\sqrt{\frac{5}{\pi}}\left(
       \begin{array}{*{3}{c}}
     1 & 0 & 0\\
     0 & 1 & 0\\
     0 & 0 & -2
       \end{array}\right),\\
{\mathscr Y}^{2\pm2}_{ij}\label{eq:y22tensor}
    &=&
       \frac{1}{4}\sqrt{\frac{15}{2\pi}}\left(
       \begin{array}{*{3}{c}}
     1 & \pm i & 0\\
     \pm i & -1 & 0\\
     0 & 0 & 0
       \end{array}\right).
\end{eqnarray}

\section{Time-dependent perturbation equations}
\label{sec:pt}

In Regge-Wheeler gauge, the metric perturbation for a given
even-parity $(\ell,m,\omega)$ mode depends on the three radial
functions $H$, $H_{1}$, and $K$. In this appendix, we introduce the
Zerilli function $\mathbb{Z}$ and the Regge-Wheeler function
$\mathbb{X}$ and describe how we obtain the radial functions from
them. The description here will hold for a general $(\ell, m,
\omega)$, while the results derived in Sec.~(\ref{sec:dynamic}) rely
on the special case when $(\ell, m, \omega)=(2, 2, 2\Omega)$.

The original Zerilli's master function is defined implicitly through
its relation with the two radial functions $H_1$ and $K$ [Eqs.~(13)
and (14) of Ref.~\cite{zerilli1} with ${R_{LM}}^{(e)}$ replaced by
$\mathbb{Z}$]:
\begin{eqnarray}
H_1
&=& -i\omega
    \frac{\lambda r^2 - 3\lambda M r -3M^2}
    {(r-2M)(\lambda r + 3M)} \mathbb{Z} -
    i\omega r\frac{d\mathbb{Z}}{dr},
    \label{eq:H1Zerilli}\\
K &=&
    \frac{\lambda(\lambda+1)r^2+3\lambda Mr+6M^2}
    {r^2(\lambda r + 3M)} \mathbb{Z} +
    \frac{d\mathbb{Z}}{dr^*},
    \label{eq:KZerilli}
\end{eqnarray}
where
\begin{equation}
\lambda \equiv \frac{1}{2}(\ell -1)(\ell +2). \nonumber
\end{equation}
Using the algebraic relationship (Eq.~(10) of
Ref.~\cite{edelstein1})
\begin{eqnarray}
\left(\frac{3M}{r}+\lambda\right)H
&=&
    \left[i\omega r -\frac{i(\lambda+1)M}{\omega r^2}\right] H_1 +
    \nonumber \\
& &
    \left(\lambda + \frac{M}{r} -\frac{M^2/r^2
        + \omega^2 r^2}{1-2M/r}\right) K
    \nonumber
\end{eqnarray}
one can obtain $H$ in terms of the Zerilli function
\begin{eqnarray}
\label{eq:HfromZ}
H =
    \left[\frac{\omega^2r^2}{2M-r} +
        \frac{s_1}{r^2 (3M+\lambda r)^2}\right] \mathbb{Z} +
    s_2 \frac{d\mathbb{Z}}{dr},
\end{eqnarray}
in which
\begin{eqnarray}
s_1 &=&
    {9M^2(M+\lambda r)} + {\lambda^2 r^2[3M+(\lambda+1)r]},
    \nonumber \\
s_2 &=&
    \frac{-3M^2 - 3\lambda Mr + \lambda r^2}{r(3M+\lambda r)}
    \nonumber.
\end{eqnarray}
The Zerilli function obeys the wave equation (Eqs.~(18) and (19) of
Ref.~\cite{zerilli1}):
\begin{eqnarray}
\label{eq:zerilli}
\left[\frac{d^{2}}{dr^{*2}} + \omega^2 - V(r)\right]\mathbb{Z} =0,
\end{eqnarray}
in which the potential term is given by
\begin{eqnarray}
V(r)
&=&
    \frac{2(r-2M)}{r^4(\lambda r+3M)^2} \Big[
    \lambda^2(\lambda+1) r^3 + \nonumber\\
& &
    \qquad\qquad\qquad\quad
    3\lambda^2 M r^2 + 9\lambda M^2 r + 9 M^3 \Big].
    \nonumber
\end{eqnarray}

The odd-parity master function, the Regge-Wheeler function, is
defined in Eq.~(23) of Ref.~\cite{regge1} (and is called $Q$ in Regge
and Wheeler's notation). It obeys the differential equation (Eq.~(7)
and of Ref.~\cite{edelstein1}):
\begin{equation}
\label{eq:rw}
\left[\frac{d^{2}}{dr^{*2}} + \omega^2 - \Big(1-\frac{2M}{r}\Big)
\Big(\frac{\ell(\ell+1)}{r^2}-\frac{6M}{r^3}\Big)\right]
\mathbb{X}=0.
\end{equation}
The connection between the Regge-Wheeler and Zerilli functions was
first found by Chandrasekhar and is listed, e.g., in Eq.~(152) of
Ch.~4 of Ref.~\cite{chandraBook}:
\begin{eqnarray}
\label{eq:Chandra}
\Big[\lambda(\lambda+1)-3iM\omega\Big]\mathbb{Z}
&=&
    \left[\lambda(\lambda+1) +
        \frac{9M^2(r-2M)}{r^2(\lambda r+3M)}\right]
    \mathbb{X}\nonumber\\
&+&
    3 M \left(1-\frac{2M}{r}\right) \frac{d\mathbb{X}}{dr}.
\end{eqnarray}
This completes our metric reconstruction scheme from the
Regge-Wheeler function. We are now ready to evaluate the radial
metric perturbation functions \(H\), \(H_1\), and \(K\) for the
$(\ell, m)=(2,2)$ mode of the perturbations. Expanding $X^H$ [given
in Eq.~(\ref{eq:xh1epsilon}) in powers of $\tilde{r}\equiv r/2M$ to
first order in $\varepsilon \equiv 2 M \omega $, we obtain
\begin{equation}
\label{eq:xHapp}
X^H =
    \EuScript{A}\left[
        \Big(1+\frac{13}{12}i\varepsilon\Big)\tilde{r}^{3}
      + \sum_{n=5}^{\infty} \frac{i\varepsilon}{n\tilde{r}^{n-3}} +
        O(\varepsilon^2)\right].
\end{equation}
Here ${\EuScript A}$ is an overall scaling factor (Eq.
(\ref{eq:Aval}) in Sec.~\ref{sec:tdmoment}). While the summation can
be rewritten as a closed-form expression, we prefer to stay in the
series notation, since our interest is in reading various powers of
$r$ in the resulting first-order tidal field. Equation
(\ref{eq:xHapp}) is the value of the Regge-Wheeler function in the
LARF; inserting it into Eq.~(\ref{eq:Chandra}) yields the expression
for $\mathbb{Z}$ in the LARF [We shall suppress
``$O(\varepsilon^2)$" hereafter]:
\begin{eqnarray}
\mathbb{Z}
&=&
   \EuScript{A}\Big(1+\frac{4i\varepsilon}{3}\Big)\Big[
        \tilde{r}^3 + \frac{3\tilde{r}^2}{4} -
        \frac{9\tilde{r}}{16} - \frac{21}{64} +
        \frac{63}{256\tilde{r}} \Big]\nonumber \\
&+&
    \EuScript{A}\Big[
        \frac{-945-236i\varepsilon}{5120\tilde{r}^2} +
        \frac{8505+15436i\varepsilon}{61440\tilde{r}^3} +
        O\Big(\frac{1}{\tilde{r}^4}\Big)
    \Big] \nonumber
\end{eqnarray}
Inserting $\mathbb{Z}$ into Eqs.~(\ref{eq:H1Zerilli}),
(\ref{eq:KZerilli}) and ({\ref{eq:HfromZ}}) yields $H_{1}$, $K$ and
$H$. Expanded in powers of $\tilde{r}$ and to first order in
$\varepsilon$, these radial functions are given by
\begin{eqnarray}
\label{eq:HKres}
H &=&
    \frac{\EuScript{A}}{M}\Big[(3+4i\varepsilon)
        (\tilde{r}^2 - \tilde{r})+
        \frac{i\varepsilon}{10\tilde{r}^3} +
        \frac{3i\varepsilon}{20\tilde{r}^4}\Big] +
        O\big(\tilde{r}^{-5}\big) \nonumber\\
H_1 &=&
    \frac{i\EuScript{A}\varepsilon}{4M}
    \Big[-8\tilde{r}^3 - {2}\tilde{r}^2 + 4\tilde{r}
        + 1
    \nonumber \\
& &
    \qquad\quad+\
    \tilde{r}^{-1}+\tilde{r}^{-2}+\tilde{r}^{-3}+\tilde{r}^{-4}\Big]
    + O(\tilde{r}^{-5}) \nonumber\\
K
&=&
    \frac{\EuScript{A}}{M}\Big[(3+4i\varepsilon)
        \Big(\tilde{r}^2 - \frac{1}{2}\Big) +
        \frac{i\varepsilon}{10\tilde{r}^3}
    \nonumber \\
& & \qquad\quad+\
        \frac{i\varepsilon}{8\tilde{r}^4} +
        \frac{9i\varepsilon}{70\tilde{r}^5}+
        \frac{i\varepsilon}{8\tilde{r}^6}
        \Big]
    + O(\tilde{r}^{-7})
\end{eqnarray}

\section{Time-independent perturbation equations}
\label{sec:apStatic} As is evident from the time-dependent
perturbation theory, as $\omega \rightarrow 0$, $H_{1}$ goes to
zero. In the static case, then, there are only two radial functions,
$\hat{H}$ and $\hat{K}$ (where the hats signify that they represent
the time-independent perturbations). Specializing to the
axisymmetric case, the metric perturbation is
\begin{eqnarray}
h^{(\ell0)}_{ab}
 =
    \text{Diag}\Big[\left(1-{2M/r}\right){\hat H},\
     \frac{{\hat H}}{1-2M/r}\ ,\quad\Big. \nonumber\\
    \Big. r^2\hat{K},\ r^2\hat{K}\sin^2\theta\Big]
    \times Y^{\ell0}(\theta,\phi).
\nonumber
\end{eqnarray}
The linearized Einstein equations governing $\hat{H}$ and $\hat{K}$
are given in Eqs.~(9d) and (9e) of Ref.~\cite{edelstein1} with
$H_1=0$ and $\omega=0$ ($k=0$ in Edelstein and Vishveshwara's
notation):
\begin{eqnarray}
\frac{d\hat{K}}{dr}
&=&
    \frac{d\hat{H}}{dr} +
    \frac{2M}{r^2}\left(1-\frac{2M}{r}\right)^{-1}\hat{H},
    \label{eq:dKdr1}\\
\frac{2M}{r^2}
\frac{d\hat{K}}{dr}
&=&
    \left(1-\frac{2M}{r}\right)\frac{d^2\hat{H}}{dr^2} +
    \frac{2}{r}\frac{d\hat{H}}{dr} -
    \frac{\ell(\ell+1)}{r^2} \hat{H}.
    \label{eq:dKdr2}
    \nonumber \\
\end{eqnarray}
Eliminating $d\hat{K}/{dr}$ from these two equations, we can then
write a single second-order differential equation for $H$ in terms
of the variable $z\equiv r/M-1$ (same as Eq.~(4.9) of
Ref.~\cite{hartle1}):
\begin{eqnarray}
(1-z^2)\frac{d^2\hat{H}}{dz^2} -2z\frac{d\hat{H}}{dz}+
\left[\ell(\ell+1)-\frac{4}{1-z^2}\right]\hat{H} =0. \nonumber
\end{eqnarray}
This takes a form of the associated Legendre differential equation.
The general solution for $\hat{H}$ is therefore
\begin{eqnarray}
\hat{H}=\alpha_\ell P^2_\ell(r/M-1) + \beta_\ell Q^2_\ell(r/M-1).
\end{eqnarray}
With the general solution for $\hat{H}$, we can integrate
Eq.~(\ref{eq:dKdr1}) or (\ref{eq:dKdr2}) to find $\hat{K}$. For
$\ell=2$, we have
\begin{eqnarray}
{\hat K}_2(r)
&=&
    \left[
      \alpha_2 P^{1}_{2}\left(r/M-1 \right)
     +\beta_2  Q^{1}_{2}\left(r/M-1 \right)
    \right] \nonumber \\
& &
    \times\frac{2M}{\sqrt{r(r-2M)}} + {\hat H}_2(r) .
\end{eqnarray}

\section{Newman-Penrose Formalism}\label{sec:npAp}
In this appendix, we summarize some equations of the Newman-Penrose
formalism for our choice of tetrad. \emph{In this Appendix and in
Sec.~\ref{sec:phase} only}, we use ingoing Eddington-Finkelstein
coordinates \((\tilde{V},r,\theta,\phi)\) and a \((+~-~-~-)\)
signature of the metric.
\subsection{Newman-Penrose quantities for Schwarzschild}

We adopt the Hartle-Hawking null tetrad, which is given by
Eqs.~(4.2) of Ref.~\cite{hartle2}, together with the normalization
conditions \(\ell^{\mu}n_\mu = 1\) and \(m^\mu \bar{m}_\mu = -1\).
The tetrad vectors have components [using the notation \(e^{\mu} =
(e^{\tilde{V}},e^r,e^\theta,e^\phi)\)]
\begin{subequations}
\begin{eqnarray}
\ell^{\mu}
&=&  \left(1,\frac{1}{2}-\frac{M}{r},0,0 \right)\\
n^{\mu}
& =&
    (0,-1,0,0)\\
m^{\mu}
&=&
    \left(0,0,\frac{1}{\sqrt{2}r},
    \frac{i}{\sqrt{2}r\sin\theta}\right)\\
\bar{m}^{\mu}
&=&
    \left(0,0,\frac{1}{\sqrt{2}r},
    -\frac{i}{\sqrt{2}r\sin\theta}\right).
\end{eqnarray}
\end{subequations}
Note that throughout this Appendix, an overbar denotes complex
conjugation.

From these basis vectors, we define the direction derivatives
\begin{equation}
D = \ell^\mu \partial_\mu\mbox{, }
\Delta = n^{\mu} \partial_\mu \mbox{, }
\delta = m^{\mu} \partial_\mu \mbox{ and }
\bar{\delta} = \bar{m}^\mu \partial_\mu.\label{eq:dirderiv}
\end{equation}

Our conventions for the spin coefficients follow Ref.~\cite{hartle2}
[specifically, Eqs.~(2.2) and (2.3)]. The spin coefficients are
defined by
\begin{subequations}
\begin{eqnarray}\label{eq:spinco}
\kappa & = & \ell_{\mu;\nu} m^\mu \ell^\nu\\
\pi & = & -n_{\mu;\nu}\bar{m}^{\mu} \ell^\nu\\
\rho & = & \ell_{\mu;\nu} m^\mu \bar{m}^\nu\\
\mu & = & -n_{\mu;\nu} \bar{m}^\mu m^\nu\\
\label{eq:spinDefs}
\sigma & = & \ell_{\mu;\nu} m^{\mu} m^\nu\\
\lambda & = & -n_{\mu;\nu} \bar{m}^{\mu} \bar{m}^\nu\\
\epsilon & = & \frac{1}{2}\left(\ell_{\mu;\nu}n^\mu \ell^\nu -
            m_{\mu;\nu} \bar{m}^\mu \ell^\nu \right).\\
\alpha & = & \frac{1}{2}\left(\ell_{\mu;\nu}n^\mu \bar{m}^\nu -
            m_{\mu;\nu} \bar{m}^\mu \bar{m}^\nu \right).\\
\beta & = & \frac{1}{2}\left(\ell_{\mu;\nu}n^\mu m^\nu -
            m_{\mu;\nu} \bar{m}^\mu m^\nu \right).\label{eq:spincoend}
\end{eqnarray}
\end{subequations}

The spin coefficients for the Schwarzschild spacetime are
\begin{subequations}
\begin{eqnarray}
\kappa & = & \sigma = \lambda = \nu = \tau = \pi = \gamma = 0\\
\epsilon & = & \frac{M}{2 r^2}\\
\rho & = & -\frac{r-2M}{2 r^2}\\
\mu & = & -\frac{1}{r}\\
\alpha & = & - \beta = -\frac{1}{2\sqrt{2}r\tan\theta}.
\end{eqnarray}\end{subequations}

Because we are only interested in vacuum regions of spacetime, the
Riemann and Weyl tensors are interchangeable. The Weyl components
are defined in vacuum by
\begin{subequations}
\begin{eqnarray}\label{eq:psi0def}
\Psi_0 & = &
        -R_{\alpha\beta\gamma\delta} \ell^\alpha
        m^\beta \ell^\gamma m^\delta\\
\Psi_1 & = &
        -R_{\alpha\beta\gamma\delta} \ell^\alpha
        n^\beta \ell^\gamma m^\delta\\
\Psi_2 & = &
        -\frac{1}{2} R_{\alpha\beta\gamma\delta}
        \left( \ell^\alpha n^\beta \ell^\gamma n^\delta +
        \ell^\alpha n^\beta m^\gamma \bar{m}^\delta \right)\\
\Psi_3 & = &
        -R_{\alpha\beta\gamma\delta}
        \ell^\alpha n^\beta \bar{m}^\gamma n^\delta\\
\Psi_4 & = &
        -R_{\alpha\beta\gamma\delta}
        n^\alpha \bar{m}^\beta n^\gamma \bar{m}^\delta.\label{eq:psi4def}
\end{eqnarray}
\end{subequations}
Their values for the Schwarzschild spacetime are
\begin{subequations}
\begin{eqnarray}
\Psi_{0} & = & \Psi_{1} = \Psi_{3} = \Psi_{4} = 0,\\
\Psi_{2} & = & -\frac{M}{r^3}.
\end{eqnarray}
\end{subequations}

The Ricci identities are
\begin{equation}
\left(\nabla_\mu \nabla_\nu - \nabla_\nu \nabla_\mu \right) e_\gamma
= R_{\sigma\gamma\mu\nu} e^{\sigma}.
\end{equation}
Inserting the null tetrad vectors for $e^\sigma$ and projecting
along the tetrad yields the Ricci identities in Newman-Penrose
notation. One of these equations is, in our tetrad and evaluated on
the horizon,
\begin{equation}\label{eq:tideAp}
D\sigma^{(1)} - 2\epsilon\sigma^{(1)} =
\partial_{\tilde{V}}\sigma^{(1)} - 2 \epsilon \sigma^{(1)} =
\Psi_0^{(1)}.
\end{equation}
(Note that we have used the fact that \(\sigma\) and \(\Psi_0\)
vanish for Schwarzschild.) This is the \emph{tidal force equation}; it relates \(\Psi_0^{(1)}\), which is physically equivalent to the tangential-tangentail component of the perturbing tidal field on the horizon, to \(\sigma\), which is physically equivalent to the shear of the instantaneous horizon.

The shape of the perturbed instantaneous horizon is determined by its two-dimensional scalar curvature \(\mathscr{R} + \mathscr{R}^{(1)}\) where \(\mathscr{R}\) is the curvature of the unperturbed horizon. According to the tidal force equation (\ref{eq:tideAp}), \(\Psi_0^{(1)}\) drives the shear, which is the ``rate of change of the shape'' of the horizon as measured by fiducial observers on the horizon \cite{MPBook}. Thus, it is not surprising that \(\mathscr{R}^{(1)}\) can be computed directly from \(\Psi_0^{(1)}\). Hartle \cite{hartle2} has derived \cite{hartle2} the explicit formula, a consequence of Gauss' relation \cite{gaussRel}, in the Newman-Penrose formalism with the present choice of coordinates and tetrad:
\begin{equation}\label{eq:shapeFromTide}
\mathscr{R}^{(1)} = -4 \text{Im}\left[\frac{(\bar{\delta}+2\pi-2\alpha)(\bar{\delta}+\pi-4\alpha)+2\epsilon\lambda}{\omega(i\omega+2\epsilon)}\right] \Psi_0^{(1)}
\end{equation} where \(\omega\) is the frequency of the perturbation. When a Schwarzschild black hole is perturbed by a distant moon in a slow, circular, orbit with angular velocity \(\Omega\), then \(\omega=2\Omega\).

\subsection{Newman-Penrose quantities for Kerr}
Finally, to facilitate our comparison to Hartle's results, we here list the relevant Newman-Penrose quantities for the Kerr spacetime [Eq.~(\ref{eq:mKerrHH})] using Hartle's choice \cite{hartle2} of coordinates and tetrad. In the limit \(a=0\), Hartle's tetrad and spin coefficients reduce to those listed in the previous subsection.

The null tetrad vectors [using the notation \(e^\mu=(e^{\tilde{V}},e^r,e^\theta,e^{\tilde{\phi}})\)] are
\begin{subequations}
\begin{eqnarray}\label{eq:lHH}
\ell^\mu & = & \left(1,\frac{r^2-2 M r + a^2}{2\left(r^2+a^2\right)},0,\frac{a}{r^2+a^2}\right) \\
n^\mu & = & \left(0,-\frac{2(a^2+r^2)}{2 r^2 +a^2 + a^2 \cos2\theta},0,0\right)\nonumber \\
& &+ \frac{-a^2+a^2 \cos2\theta}{2\left(a^2+2r^2+a^2 \cos2\theta\right)} \ell^\mu\nonumber\\ & &  + \frac{-a\sin\theta}{\sqrt{2}\left(i r + a \cos\theta\right)} m^\mu \nonumber\\ & & + \frac{-a\sin\theta}{\sqrt{2}\left(-i r + a \cos\theta\right)} \bar{m}^\mu \\
m^\mu & = & \left(0, -\frac{a\sin\theta \left(r^2-2 M r + a^2\right)}{2\sqrt{2}\left(r^2+a^2\right)\left(-i r + a \cos\theta\right)},\right.\nonumber\\
& & \left. \frac{1}{\sqrt{2}\left(r+i a \cos\theta\right)},\frac{\left(i r + a \cos\theta\right)\csc\theta}{\sqrt{2}\left(r^2+a^2\right)}\right)\nonumber\\
& & \\
\bar{m}^\mu & = & \left(0, -\frac{a\sin\theta \left(r^2-2 M r + a^2\right)}{2\sqrt{2}\left(r^2+a^2\right)\left(i r + a \cos\theta\right)},\right.\nonumber\\
& & \left. \frac{1}{\sqrt{2}\left(r-i a \cos\theta\right)},\frac{\left(-i r + a \cos\theta\right)\csc\theta}{\sqrt{2}\left(r^2+a^2\right)}\right)\label{eq:mBarHH}\nonumber\\
& & 
\end{eqnarray}
\end{subequations} Then, one can compute the spin coefficients for this tetrad from Eqs.~(\ref{eq:spinco})--(\ref{eq:spincoend}): 
\begin{subequations}
\begin{eqnarray}
\kappa & = & \sigma = 0\\
\lambda & = & O(a^2)\\
\nu & = & O(a^2)\\
\tau & = & \frac{-i(2M+r) \sin\theta a}{2\sqrt{2}r^3}\\
\pi & = & \frac{i(4M+r) \sin\theta a}{2\sqrt{2}r^3}\\
\gamma & = & \frac{-i \cos\theta a}{2 r^2} + O(a^2)\\
\epsilon & = & \frac{M}{2} \frac{r^2-a^2}{\left(r^2+a^2\right)^2}= \frac{M}{2r^2}+O(a^2)
\end{eqnarray}
\begin{eqnarray}
\rho & = & -\frac{r-2M}{2 r^2} - \frac{i(r-2M)\cos\theta a}{2 r^3} \nonumber\\ & & + O(a^2) \\
\mu  & = & -\frac{1}{r} + O(a^2) \\
\alpha & = & \frac{-\cot\theta}{2\sqrt{2}r} - \frac{i[-3M+(2M+2r)\cos 2\theta]a}{4\sqrt{2}r^3 \sin\theta}\nonumber\\ & & + O(a^2)\\
\beta & = &  \frac{\cot\theta}{2\sqrt{2}r} - \frac{i[M+r+(r-M)\cos 2\theta] a}{4\sqrt{2} r^3 \sin\theta} \nonumber\\& & + O(a^2).
\end{eqnarray} 
\end{subequations} 

The directional derivatives are then given by Eq.~(\ref{eq:dirderiv}).

Using the Kerr metric [Eq.~(\ref{eq:mKerrHH})] and Hartle's choice for the tetrad [Eqs.~(\ref{eq:lHH})--(\ref{eq:mBarHH})], one can compute the curvature for Kerr and read off the curvature scalars via Eqs.~(\ref{eq:psi0def})--(\ref{eq:psi4def}):

\begin{subequations}
\begin{eqnarray}\label{eq:psiVanishKerr}
\Psi_0 & = & \Psi_1 = 0\\
\Psi_2 & = & -\frac{M}{\left(r-i a \cos\theta\right)^3}\\
\Psi_3 & = & -\frac{3 i a M \sin\theta}{\sqrt{2}\left(r-i a \cos\theta\right)^4}\\
\Psi_4 & = & \frac{3 i a^2 M \sin^2\theta}{(i r + a \cos\theta)^5}.
\end{eqnarray}
\end{subequations}

The tidal force equation (\ref{eq:tideAp}) relates \(\Psi_0^{(1)}\) to \(\sigma^{(1)}\). The correction to the scalar curvature of the horizon, \(\mathscr{R}^{(1)}\), is given by Eq.~(\ref{eq:shapeFromTide}). 

For a stationary moon perturbing a slowly rotating Kerr black hole, the frequency of the perturbation is \(\omega = -2\Omega_H = -8 M^2 \Omega_H \).


\begin{thebibliography}{34}
\bibitem[{AL()}]{AL}
\bibinfo{note}{We follow the same notation as in Ref. \cite{thorne1}}.
\end{thebibliography}

\end{document}